\documentclass{aa}  
\usepackage{graphicx}
\usepackage{txfonts}
\usepackage{xcolor}
\usepackage{verbatim}

\begin{document}

   \title{Edge collapse and subsequent longitudinal accretion in Filament S242}


   \author{Lixia Yuan\inst{1,2,3}\and 
   Guang-Xing Li\inst{4}\and
   Ming Zhu\inst{1,2,3}\and
   Tie Liu\inst{5}\and 
   Ke Wang\inst{6}\and 
   Xunchuan Liu\inst{7, 6}\and 
   Kee-Tae Kim\inst{8, 9}\and \\
   Ken'ichi Tatematsu\inst{10, 11}\and 
   Jinghua Yuan\inst{1}\and 
   Yuefang Wu\inst{7, 6}
   }

   \institute{National Astronomical Observatories, Chinese Academy of Sciences,
20A Datun Road, Chaoyang District, Beijing 100012, China \\
              \email{lxyuan@nao.cas.cn, mz@nao.cas.cn}
         \and
          University of Chinese Academy of Sciences,
100049, Beijing, PR China \\
        \and
          Key Laboratory of FAST, NAOC, Chinese Academy of Sciences, Beijing 100012, PR China \\
          \and
          South-Western Institute for Astronomy Research, Yunnan University, Kunming, 650500 Yunnan, PR China \\   \email{gxli@ynu.edu.cn}
          \and
          Key Laboratory for Research in Galaxies and Cosmology, Shanghai Astronomical Observatory, Chinese Academy of Sciences,
80 Nandan Road, Shanghai 200030, China\\
          \and
          Kavli Institute for Astronomy and Astrophysics, Peking University, 5 Yiheyuan Road, Haidian District, Beijing 100871, PR China \\
          \and
          Department of Astronomy, Peking University, 100871, Beijing, PR China \\
          \and
          Korea Astronomy and Space Science Institute, 776
Daedeokdae-ro, Yuseong-gu, Daejeon 34055, Republic of Korea \\ 
          \and
          University of Science and Technology, Korea (UST), 217 Gajeong-ro, Yuseong-gu, Daejeon 34113, Republic of Korea \\
          \and
          Nobeyama Radio Observatory, National Astronomical Observatory of Japan, National Institutes of Natural Sciences, 462-2 Nobeyama, Minamimaki, Minamisaku, Nagano 384-1305, Japan \\
          \and
          Department of Astronomical Science, SOKENDAI (The Graduate University for Advanced Studies), 2-21-1 Osawa, Mitaka, Tokyo 181-8588, Japan \\
           }

  \date{Received September 3th, 2019 ; accepted March 31th, 2020}

 
  \abstract
  {Filament S242 is  25 pc long with massive clumps and YSO clusters concentrated in its end regions; it is considered a good example of edge collapse. We mapped this filament in the $^{12}$CO(1-0) and $^{13}$CO(1-0) lines. A large-scale velocity gradient along filament S242 has been detected; the relative velocity between the two end-clumps is $\sim$ 3 km s$^{-1}$, indicating an approaching motion between them. These signatures are consistent with the filament S242 being formed through the collapse of a single elongated entity, where an effect known as ``gravitational focusing'' drives the ends of the filament to collapse (edge collapse). Based on this picture, we estimate a collapse timescale of $\sim$ 4.2 Myr, which is the time needed for a finite and elongated entity evolving to the observed filament S242. For the whole filament, we find that increases in surface densities lead to increases in velocity dispersion, which can be consistently explained as the result of self-gravity. We also calculated the contribution of longitudinal collapse to the observed velocity dispersion and found it to be the dominant effect in driving the gas motion near the end-clumps. We propose that our filament S242 is formed through a two-stage collapse model, where the edge collapse of a truncated filament is followed by a stage of longitudinal accretion toward the dense end-clumps.}


\maketitle
%

\section{Introduction}
Molecular clouds exhibit a complex, irregular, and filamentary morphology \citep{Bally1987, Wiseman1998, Goldsmith2008, Busquet2013, Andre2016}. Recent dust continuum surveys, including the APEX Telescope Large Area Survey of the Galaxy (ATLASGAL; \citealt{Schuller2009}) and the Herschel infrared Galactic Plane Survey (Hi-GAL; \citealt{Molinari2010}), have revealed that filaments are ubiquitous along the Galactic plane. These filamentary structures have wide ranges of masses ($\sim$ 1 -- 10$^{5}$ M$_{\odot}$) and lengths ($\sim$ 0.1 -- 100 pc) \citep{Kirk2013, Hacar2013, Wang2015, Wang2016, Ligx2013, Ligx2016, Mattern2018, Kainulainen2013, Kainulainen2017}. Moreover, a significant fraction of gravitationally bound dense cores and protostars are embedded in filaments, which implies that filaments play an important role in the process of star formation \citep{Andre2013, Andre2014, Andre2017, Schneider2012, Contreras2016, Li2016, Liua2018, Liub2018, Lu2018, Yuanlx2019}. 

The process of filament formation and the roles of filaments in star formation are not yet well understood. 
It is hypothesized that a star-forming cloud prior collapses to lower dimensionality structures, such as sheets and filaments, the resulting layers or filaments will further fragment as a result of the gravitational instability \citep{Schneider1979, Larson1985}. Infinite isothermal sheets are perturbed to form successive and parallel filaments; furthermore, infinite isothermal filaments tend to fragment into the strings of dense cores \citep{Inutsuka1992, Curry2000, Myers2009}. For finite sheets and filaments, simulations show that gravitational focusing can result in strong density enhancements of dense gas in their edge-regions \citep{Burkert2004, Hartmann2007, Heitsch2008, Pon2011, Pon2012}. Meanwhile, far-infrared and submillimeter  observations reveal that the dense gas-clumps are mainly concentrated at the ends of filamentary clouds \citep{Hacar2013, Dewangan2017, Johnstone2017, Kainulainen2017, Ohashi2018}.

\cite{Burkert2004} suggested  that non-spherical substructures collapse on timescales longer than equal-density spherical objects. Uniform-density spheres collapse homologously \citep{Binney1987}; instead,  for structures like sheets and filaments the edge effect causes their edges to be preferentially accelerated, and then an infalling edge would sweep up the interior material \citep{Burkert2004, Pon2011, Pon2012, Clarke2015}. 

The molecular cloud S242 is a spatially elongated filamentary structure with a length of $\sim$ 25 pc at the distance of 2 kpc \citep{Yuanlx2019, Dewangan2017, Dewangan2019}. The northern end regions of S242 include two Planck Galactic cold clumps (PGCCs) G181.84+0.31 (hereafter    G181) and G182.04+0.41 (hereafter G182), as presented in Figure \ref{fig:fco} \citep{Planck2016}. The southern end region (hereafter S242-S) is associated with an HII region, which is ionized by the star BD+26 980 \citep{Hunter1990}. Recently, \cite{Dewangan2017, Dewangan2019} found that most massive clumps and YSOs clusters are located at the ends of S242 and concluded that filament S242 is a good example to study the end-dominated collapse. Some candidates, such as the Serpens filament, are velocity-coherent (trans)sonic structures and at the onset of slightly supercritical collapse \citep{Gong2018}. Our filament S242 with larger Mach number of $\sim$ 2.7 -- 4.0 is indicative of supersonic turbulent motion \citep{Dewangan2019}. In addition, clear evidence of sequential star formation in G181 are revealed, which is supposed to be caused by the edge effect in the collapse of S242 \citep{Yuanlx2019}. 

In this paper we analyze the edge collapse and subsequent longitudinal accretion in filament S242. This paper is organized as follows. Section 2 presents the TRAO observation and data reduction. Section 3 presents the results, including the basic physical and kinematical information of filament S242. In Section 4 we discuss the kinematics of end-dominated collapse and the relation between self-gravity and turbulence motion. Section 5 presents the conclusions of this study.


\section{Observations and data reduction}

We performed the on-the-fly (OTF) mapping observations at the $J$=1-0 transition lines of $^{12}$CO and $^{13}$CO for the whole molecular cloud S242 using the 14-meter Radio Telescope\footnote{The new receiver system, SEQUOIA-TRAO, is equipped with high-performance 16-pixel MMIC preamplifiers in a 4$\times$4 array, operating within the 85-115 GHz frequency range. The backend system (an FFT spectrometer) provides 4096$\times$2 channels with fine velocity resolution of $\sim$ 0.04 km $s^{-1}$(15 kHz) per channel, and full spectral bandwidth of 62.5 MHz($\sim$ 160 km $s^{-1}$ for $^{12}$CO).} of Taeduk Radio Astronomy Observatory (TRAO)\footnote{Taeduk Radio Astronomy Observatory is a branch of the Korea Astronomy and Space Science Institute (KASI) in Daejeon, South Korea.}. We  divided the molecular cloud S242 into two OTF maps (32 arcmin $\times$ 32 arcmin for each map). The FWHM beam sizes ($\theta_{B}$) for the $^{12}$CO(1-0) and $^{13}$CO(1-0) lines are 45 arcsec and 47 arcsec, respectively. The main beam efficiencies ($\eta_{B}$) for $^{12}$CO(1-0) and $^{13}$CO(1-0) lines are 0.54 and 0.51, respectively. An RMS noise level of $\sim$ 0.8 K in T$_{A*}$ for the $^{12}$CO(1-0) line and $\sim$ 0.4 K  for the $^{13}$CO(1-0) line are achieved with a spectral resolution of $\sim$ 0.04 km s$^{-1}$. The OTF data are gridded into the Class format files with a pixel size of 24 arcsec using the OTF tool at TRAO.


\section{Results}
   \begin{figure*}
   \centering
   \includegraphics[width=0.9\hsize]{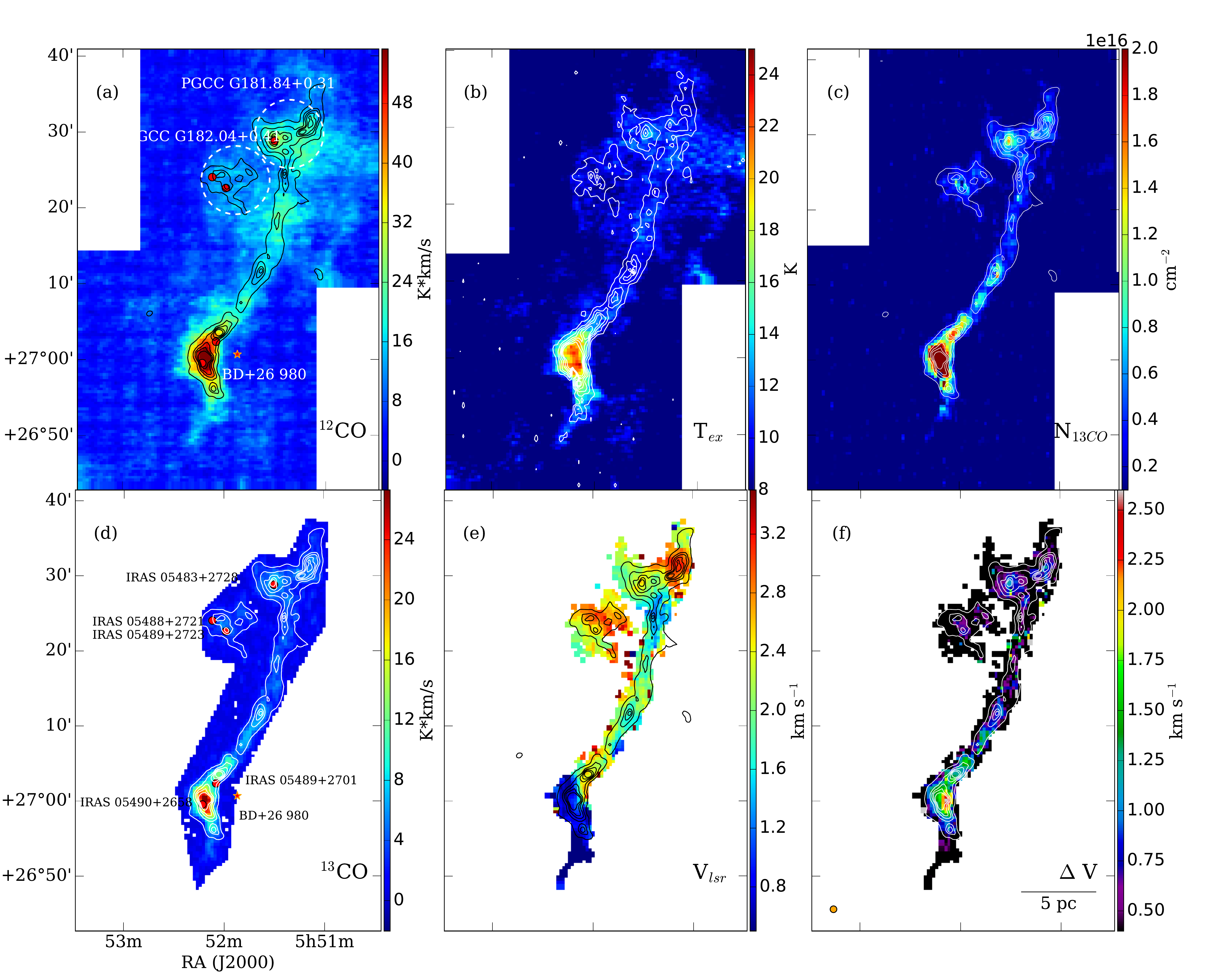}
   \caption{\textbf{Panels (a) and (d)}:  Color-scale images represent the distributions of the integrated intensity for $^{12}$CO(1-0) and $^{13}$CO(1-0) lines in the velocity range of -5 km s$^{-1}$ -- 10 km s$^{-1}$. The red circles denote the IRAS sources in the filament S242. The orange star represents the star BD+26 980, which triggers the HII region associated with the S242-S region. The white dashed circles indicate the Planck Galactic cold clumps G181.84+0.31 and G182.04+0.41. The black and white contours in all the panels, except for panel (b), display the H$_{2}$ column density distribution of S242, which derived as \cite{Dewangan2017}, ranging from 25 \%\ to 90 \% stepped by  15 \%\ of the value (2.0 $\times$ 10$^{22}$ cm$^{-2}$). \textbf{Panel (b)}:  Excitation temperature map derived from $^{12}$CO(1-0) line. The white contours display $^{13}$CO column density distribution of S242, ranging from 10 \%\ to 90 \%\ stepped by  15 \%\ of the value (2.0 $\times$ 10$^{16}$ cm$^{-2}$). \textbf{Panel (c)}:  $^{13}$CO column density map. \textbf{Panel (e)}:  Centroid velocity field of S242, which derived from the Gaussian fitting of $^{13}$CO(1-0) line. \textbf{Panel (f)}:  Distribution of FWHM velocity width for S242, which derived from the Gaussian fitting of $^{13}$CO(1-0) line. The orange circle in  panel (f) shows the effective beam size of TRAO.}
\label{fig:fco}%
\end{figure*}

\subsection{The integrated intensity of CO lines}

Figure \ref{fig:fco} shows the integrated intensity distributions of $^{12}$CO(1-0) line in Panel (a) and $^{13}$CO(1-0) line in Panel (d) in the S242 field. Only the region with CO peak intensity higher than 2 K (5 $\sigma$, an RMS noise of 0.4 K at 0.2 km s$^{-1}$ for T$_{\rm mb}$) is considered. We find that $^{13}$CO emission mainly distributes along the extended S242 and agree well with the dust emission. The CO intensity values are higher in both end regions of the S242, especially for S242-S. While the $^{12}$CO line is optically thick and could trace relatively diffuse molecular gas. We find that there is diffuse molecular gas in the gap between  G181 and G182. \cite{Dewangan2019} have presented the integrated intensity maps of $^{13}$CO(1-0), C$^{18}$O(1-0), and CS(2-1) emission, the dense gas traced by C$^{18}$O(1-0) and CS(2-1) lines are also mainly located in northern G181, G182, and S242-S regions. 

\begin{figure*}
\centering
\includegraphics[width=0.86\hsize]{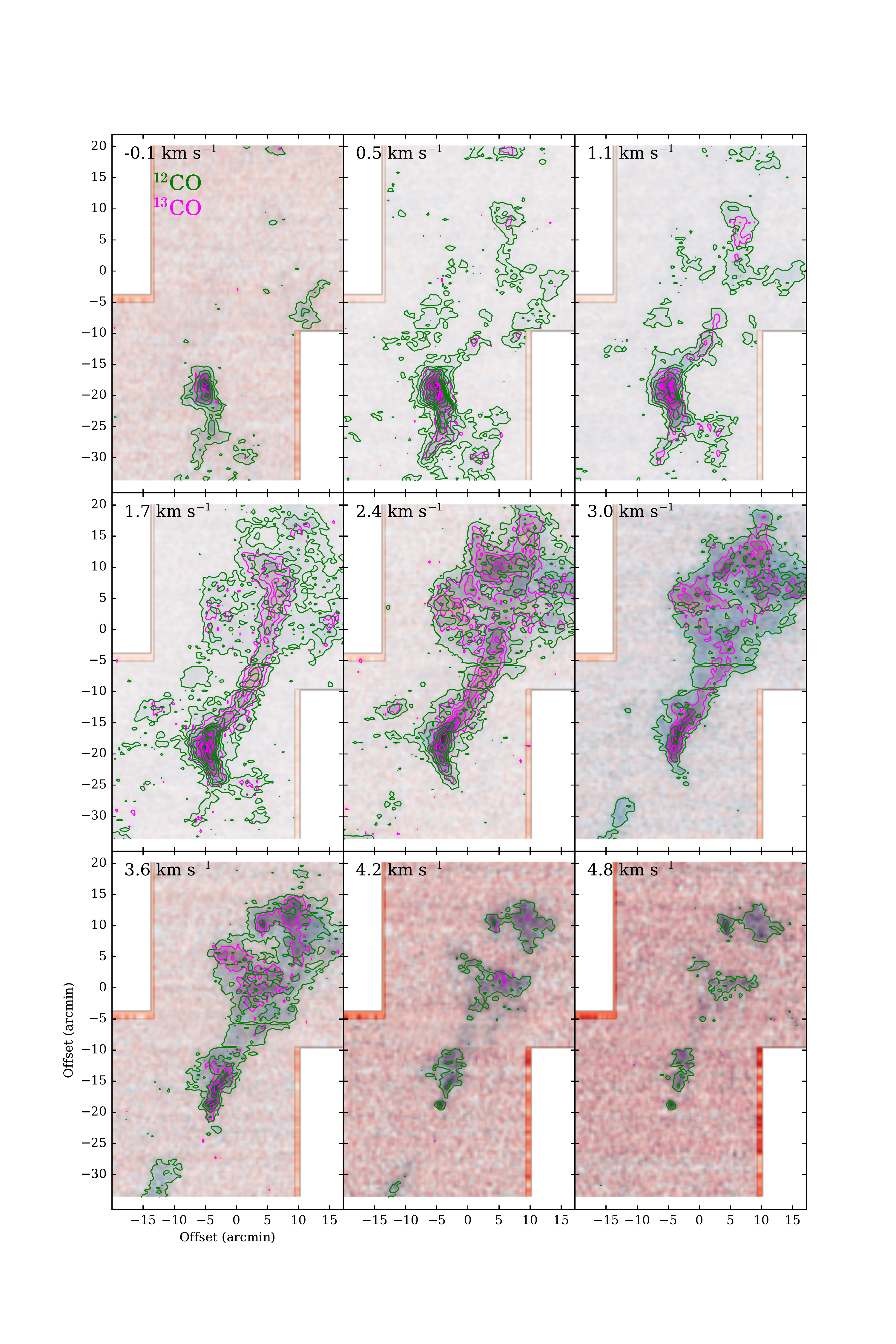}
\caption{Channel Maps of $^{12}$CO(1-0) and $^{13}$CO(1-0) lines. The green and magenta contours display the distribution of the integrated intensity $^{12}$CO(1-0) and $^{13}$CO(1-0) lines, respectively, in the interval velocity range of 0.6 km s$^{-1}$. The levels of green contours are from  6$\sigma$ ($\sim$ 1.2 K km s$^{-1}$) to the peak value by   intervals of 6$\sigma$. The levels of magenta contours are from   3$\sigma$ ($\sim$ 0.6 K km s$^{-1}$) to the peak value by   intervals of 3$\sigma$. \label{fig:fchannelmap}}
\end{figure*}

\subsection{Excitation temperature map and $^{13}$CO column densities of S242}
We estimated the excitation temperature and $^{13}$CO column density using the $^{12}$CO and $^{13}$CO line data with equations (4) -- (13) in \cite{Qian2012}. 
The resultant gas excitation temperature map of S242 is presented in  panel (b) of Figure \ref{fig:fco}, where most of the gas is in the range between 8 K and 25 K. 
Prior dust temperature estimated for S242 ranges from 10 K to 27 K \citep{Dewangan2017}, which seems to be in a good agreement with gas excitation temperature. In S242, the higher temperatures ($\sim$ 15 -- 25 K) in the S242-S region may be caused by the warm gas associated with the S242 HII region or the active star formation activities revealed in \cite{Dewangan2017}, while molecular gas away from the S242-S region is in the temperature range of about 10 K -- 14 K.

As presented in the panel (c) of Figure \ref{fig:fco}, $^{13}$CO column density(N(CO)) distributes in the range of 0.2 $\times$ 10$^{16}$ cm$^{-2}$ -- 2 $\times$ 10$^{16}$ cm$^{-2}$. N(CO) is influenced by chemical processes, such as CO depletion at high column densities \citep{Kramer1999, Caselli1999, Tafalla2002} and the CO formation and destruction at low column densities \citep{Van1988, Visser2009}. Moreover, temperature gradients in the molecular clouds also affect N(CO) by $\sim$ 30\%\ -- 70\%\ \citep{Evans2001, Pineda2010}. We find the gas distribution is very consistent  with the  H$_{2}$ column density estimated from the Herschel dust continuum emission. In addition, the densest gas-clumps are mainly concentrated on S242-S.

\subsection{Centroid velocity and velocity-width of $^{13}$CO(1-0) molecular lines}
Panels (e) and (f) of Figure \ref{fig:fco} display the distribution of the line-of-sight centroid velocity (V$_{\rm lsr}$) and FWHM velocity-width ($\Delta$V), both of which are derived by fitting a single Gaussian to the $^{13}$CO(1-0) line profile. We limited the available $^{13}$CO(1-0) line emission to regions where the intensity value are higher than $\sim$ 8 $\sigma$ (3.0 K, T$_{\rm mb}$). The fitted $^{13}$CO emission is within the velocity range of - 5 km s$^{-1}$ -- 10 km s$^{-1}$. The value of V$_{\rm lsr} $ map ranges from 0.5 km s$^{-1}$ to 3.5 km s$^{-1}$; we find that the northern end is red-shifted ( 2.5 km s$^{-1}$ -- 3.5 km s$^{-1}$), while the southern S242-S is mainly blue-shifted (0.5 km s$^{-1}$ -- 1.0 km s$^{-1}$). 
Furthermore, the large-scale centroid velocity gradients, from bluer clump at the southern end to the redder one at the northern end, may represent the gas flow along the filament or the approaching movement between the two end-clumps. However, such a gradual change in velocity is disturbed in the gas near end-clumps, which show larger local velocity-gradients of $\sim$ 1 km s$^{-1}$ pc$^{-1} $ against S242-S and $\sim$ 0.5 km s$^{-1}$ pc$^{-1} $ against the northern clump. The FWHM velocity-width distribute in a range of 0.5 km s$^{-1}$ -- 2.5 km s$^{-1}$. We find that there is an increasing velocity width (FWHM) toward the two end-clumps, especially for that in S242-S with a peak value of 2.5 km s$^{-1}$. 


\subsection{Channel maps of CO molecular lines for S242}
Figure \ref{fig:fchannelmap} presents the channel maps of $^{13}$CO(1-0) and $^{12}$CO(1-0) lines emission in the increment of 0.6 km s$^{-1}$ ranging from - 0.1 km s$^{-1}$ to 4.8 km s$^{-1}$. In the low-velocity range of -0.1 km s$^{-1}$ -- 1.1 km s$^{-1}$, the molecular gas mainly locates in the S242-S region. As the velocity increases from 1.1 km s$^{-1}$ to 2.4 km s$^{-1}$, the CO emission gradually extends from  S242-S to the northern end (G181 and G182). The gas in the northern end is mainly in the  red-shifted velocity range of 2.4 km s$^{-1}$ -- 3.6 km s$^{-1}$. Our $^{12}$CO observations reveal much more diffuse molecular gas, which was  missed by \cite{Dewangan2019}. In the red-shifted velocity range of 2.4 km s$^{-1}$ -- 3.6 km s$^{-1}$, both G181 and G182 are included in the diffuse structure of S242. 

\subsection{Filamentary structures in molecular cloud S242}
\subsubsection{Skeleton structures of Filament S242}
\begin{figure}
\centering
\includegraphics[width=\hsize]{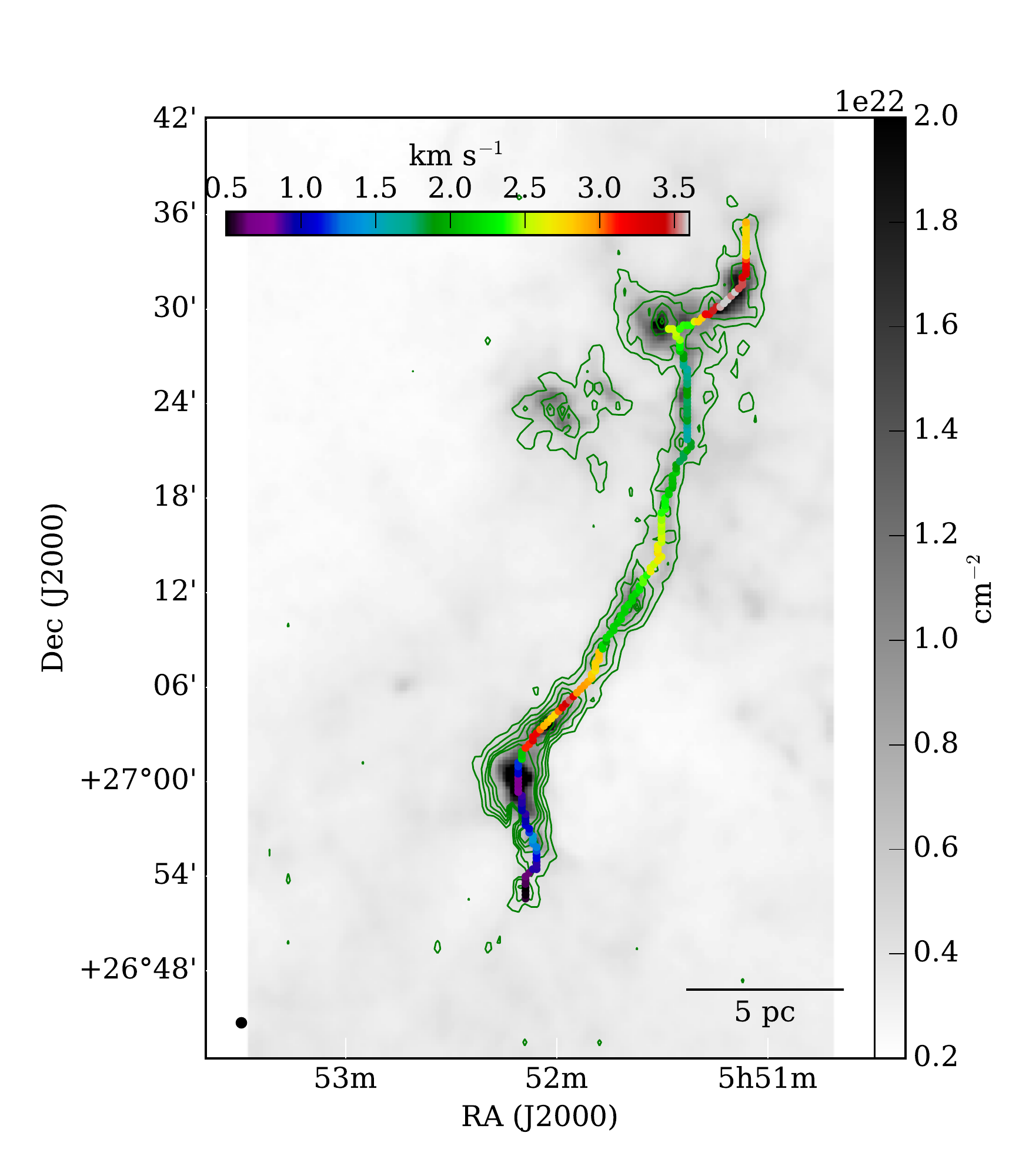}
\caption{Skeletons of the filamentary network extracted with the FilFinder, color-coded by the centroid velocities of the $^{13}$CO spectral line. The gray map is the H$_{2}$ column density map derived from Herschel continuum map (160 $\mu$m -- 500 $\mu$m) in \cite{Dewangan2017}. The overlaid green contours display $^{13}$CO column density distribution of filament S242, ranging from 15 \%\ to 90 \%\ stepped by   15 \%\ of the value (1.5 $\times$ 10$^{16}$ cm$^{-2}$). \label{fig:fskeleton}}
\end{figure}

We used the FILFINDER algorithm \citep{Koch2015} to detect the filamentary structures in H$_{2}$ column density map, which is derived by Herschel dust continuum emission (160 $\mu$m -- 500 $\mu$m) in the same way as in \cite{Dewangan2017}. The FILFINDER procedure has been successfully applied in \cite{Yuanlx2019}. In  Figure \ref{fig:fskeleton}, we plot the skeletons of the filament S242, color-coded by the centroid velocities of $^{13}$CO(1-0) line. Combing this data with  panel (e) of Figure \ref{fig:fco}, it is evident that the extended skeletons along the whole S242 filament present global velocity gradients, especially for the substructures in both end regions of S242. 
 
In addition, the width of filament S242 can be calculated by building a radial profile of H$_{2}$ column density with respect to the skeleton. As presented in Figure A.1 of the Appendix, we use RadFil, a publicly available Python package to build and fit a radial profile for filament S242 \citep{Zucker2018}. The radial profiles are built by taking radial cuts across the above skeletons extracted by FILFINDER, thereby presenting the radial structure of the filament across its entire length.  In panel (a) of Figure A.1 we present the distribution of cuts across the spine, where we determine an approximate distance between cuts of $\sim$ 0.7 pc (the parameter of samp\_int = 5 in the RadFil procedure means the cuts are $\sim$ 5 pixels apart, corresponding to $\sim$ 0.7 pc at a pixel scale of 14 arcsec and  distance of 2 kpc). In panels (b) and (c) of Figure A.1, Plummer and Gaussian functions are used to fit the entire ensemble of radial cuts, respectively. We get a best-fit  flattening radius of 0.5$\pm$0.16 pc using Plummer fitting function, a non-deconvolved FWHM-width of 0.93 pc and a deconvolved FWHW-width of 0.86 pc with the Gaussian fitting function.

\subsubsection{Main velocity component of Filament S242}
\begin{figure*}
\centering
\includegraphics[width=\hsize]{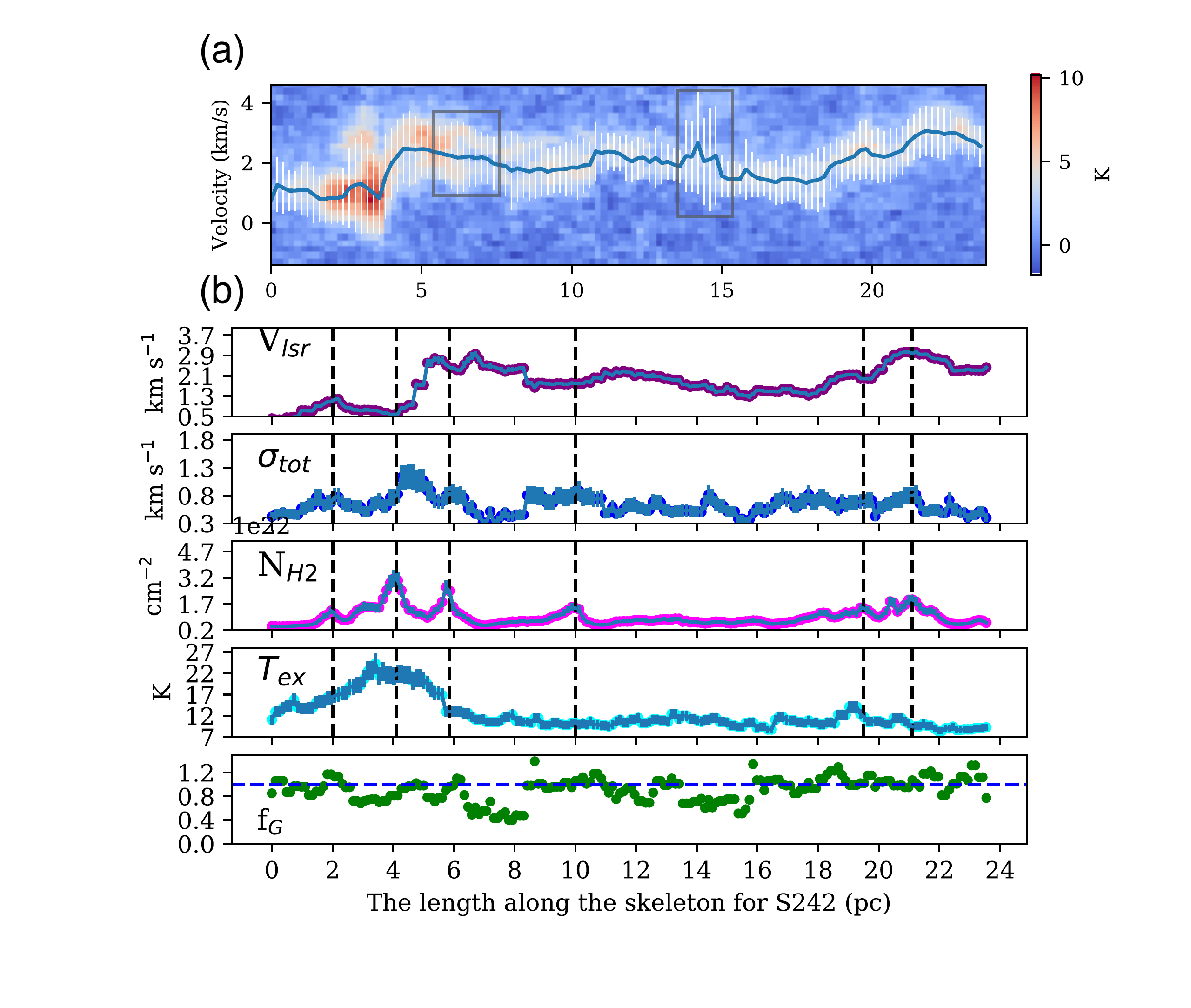}
\caption{\textbf{Panel (a)}: Position-velocity distribution along the spine of filament S242 from the southern to northern end. The dark blue curve and vertical white lines represent the fitted centroid velocity and FWHM velocity width by a single Gaussian, respectively. \textbf{Panel (b)}: Centroid velocity (V$_{\rm lsr}$), total 1D velocity dispersion ($\sigma_{\rm tot}$), H$_{2}$ column density (N$_{\rm H_{2}}$), excitation temperature (T$_{\rm ex}$), and the fractional intensity of the fitted main velocity components taken in the observed spectra (f$_{\rm G}$), distributed along the main skeleton of the filament S242. The vertical black dashed lines indicate the peaks of the H$_{2}$ column density profile; the horizontal blue dashed line represents f$_{\rm G}$ = 1. \label{fig:fskeleton_profile}}
\end{figure*}

In panel (a) of Figure \ref{fig:fskeleton_profile} we present the position-velocity diagram produced along the spine of the filament S242, as well as the centroid velocity and FWHM velocity-width obtained by fitting a single Gaussian along the spine, as shown by the dark blue curve and vertical white lines, respectively. In some regions, which are indicated as black rectangles, they cannot be fitted by a single Gaussian. The spectra of $^{13}$CO(1-0) lines and their associated fits along the spine are also shown in Appendix Figure A.3; each profile is the averaged spectrum in each magenta circle (Figure A.2) with a radius of 0.5 pc. The profiles of $^{13}$CO(1-0) lines are fitted by either one or two Gaussians. We find that the average $^{13}$CO(1-0) lines in the magenta circle of No.6 -- No.8 and No.13 -- No.15 show two peaks. The existence of multiple components is validated with the optical thin line C$^{18}$O(1-0). As presented in Figure A.3, we find the C$^{18}$O spectra are also double peaked at the locations where $^{13}$CO double peaked. These C$^{18}$O line data are from the Milky Way Imaging Scroll Painting project (MWISP), which is an unbiased Galactic plane CO survey with the 13.7 m telescope of the Purple Mountain Observatory \citep{Yang2019}.

Except for the noted structures in the length ranges of $\sim$ 6.5 pc -- 8.5 pc and $\sim$ 13.5 pc -- 15.5 pc, which show two velocity components, most regions are mainly dominated by one velocity component. Therefore, the spines are fitted with a single Gaussian or double Gaussians accordingly. Then we pick one component as the main component along filament S242. The main component is defined as the velocity component, which accounts for a larger fraction and is coherent with adjacent structures. In panel (b) of Figure \ref{fig:fskeleton_profile}, we present the fractional intensity (f$_{\rm G}$) of the main velocity component taken in the total observed velocity components, the f$_{\rm G}$ is calculated as the ratio of the velocity-integrated intensity of the fitting Gaussian for each main velocity component to that of the observed spectrum. As presented, the values of f$_{\rm G}$ in the length range of $\sim$ 6.5 pc -- 8.5 pc and $\sim$ 13.5 pc -- 15.5 pc are from 0.5 to 0.7, the values in other regions are close to 1. In addition, at each location with two velocity components, the fitted centroid velocity for the main velocity component in the $^{13}$CO(1-0) line is nearly consistent with that traced by C$^{18}$O line, as presented in Figure A.3.

\subsubsection{Longitudinal profiles of the main skeleton in Filament S242}
Panel (b) of Figure \ref{fig:fskeleton_profile} presents the profiles of fitted centroid velocity, velocity dispersion ($\sigma_{\rm tot}$), H$_{2}$ column density (N$_{\rm H_{2}}$),  and excitation temperature (T$_{\rm ex}$) for the main skeleton from the southern to northern end-point (only the eastern skeleton in G181 is included). The southern end-point is defined as the zero point of the x-axis. The x-axis value of a pixel on the skeleton represents its distance from the zero point integrated along the spine line in  Figure \ref{fig:fskeleton}. The vertical dashed black lines denote the peaks of H$_{2}$ column densities (N$_{\rm H_{2}}$) along the skeleton.

The total 1D velocity dispersion ($\sigma_{\rm tot}$) for the mean molecules is calculated using \citep{Fuller1992}\begin{equation}
\rm \sigma_{\rm tot}=\sqrt{\frac{\Delta v_{\rm obs}^{2}}{8ln(2)} + k_{\rm B}T_{\rm kin}\left(\frac{1}{\mu m_{\rm H}}- \frac{1}{m_{\rm obs}}\right)}\ ,
\end{equation}
where $\Delta v_{\rm obs}$ is the observed line width (FWHM) derived by the Gaussian fitting of the C$^{13}$O lines; $k_{\rm B}$ is the Boltzmann constant; $\mu$=2.33, the atomic weight of the mean molecule; m$_{\rm H}$ is the mass of a hydrogen atom; and m$_{\rm obs}$ is the mass of the observed molecule (29 a.m.u for C$^{13}$O), T$_{\rm kin}$ is the kinetic temperature of gas, which is assumed to be the value of excitation temperature. The uncertainties for $\sigma_{\rm tot}$ are mainly contributed from the measured FWHM of $\sim$ 5$\%$, an estimated excitation temperature of $\sim$ 10 $\%$ and optical depth of $\sim$ 20$\%$ \citep{Hacar2016}. In addition, \cite{Dewangan2019} derived the Mach number of  $\sim$ 2.7 -- 4.0 for the subregions in filament S242 based on the C$^{18}$O spectra, suggesting that all the subregions are supersonic. These subregions are marked in Figure 6 of \cite{Dewangan2019} and cover our skeleton structures for the filament S242.

\begin{figure*}
\centering
\includegraphics[width=\hsize]{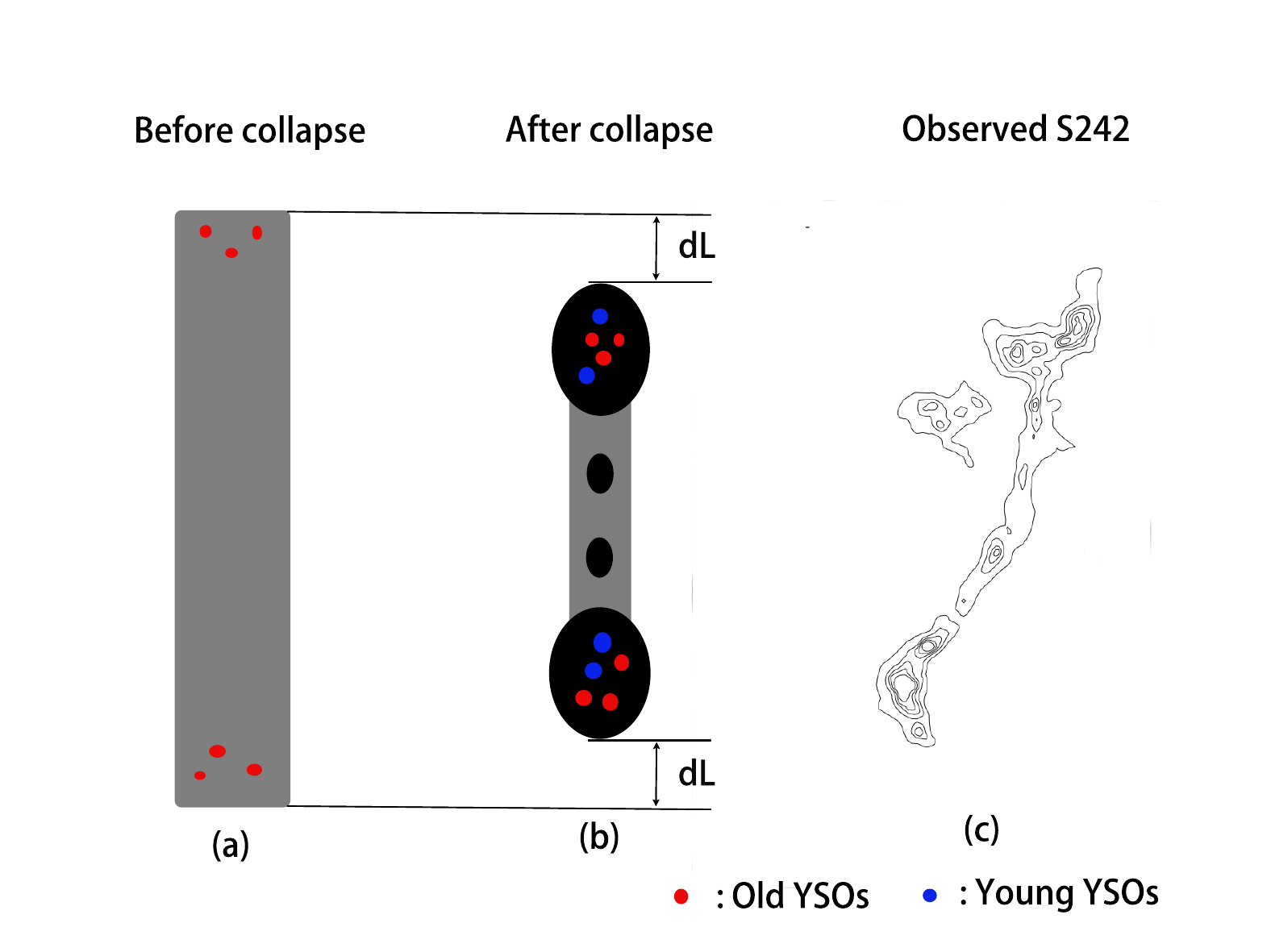}
\caption{Sketch of the edge-collapse process of the filament S242. \label{fig:fsketch}}
\end{figure*}

\begin{figure}
\centering
\includegraphics[width=\hsize]{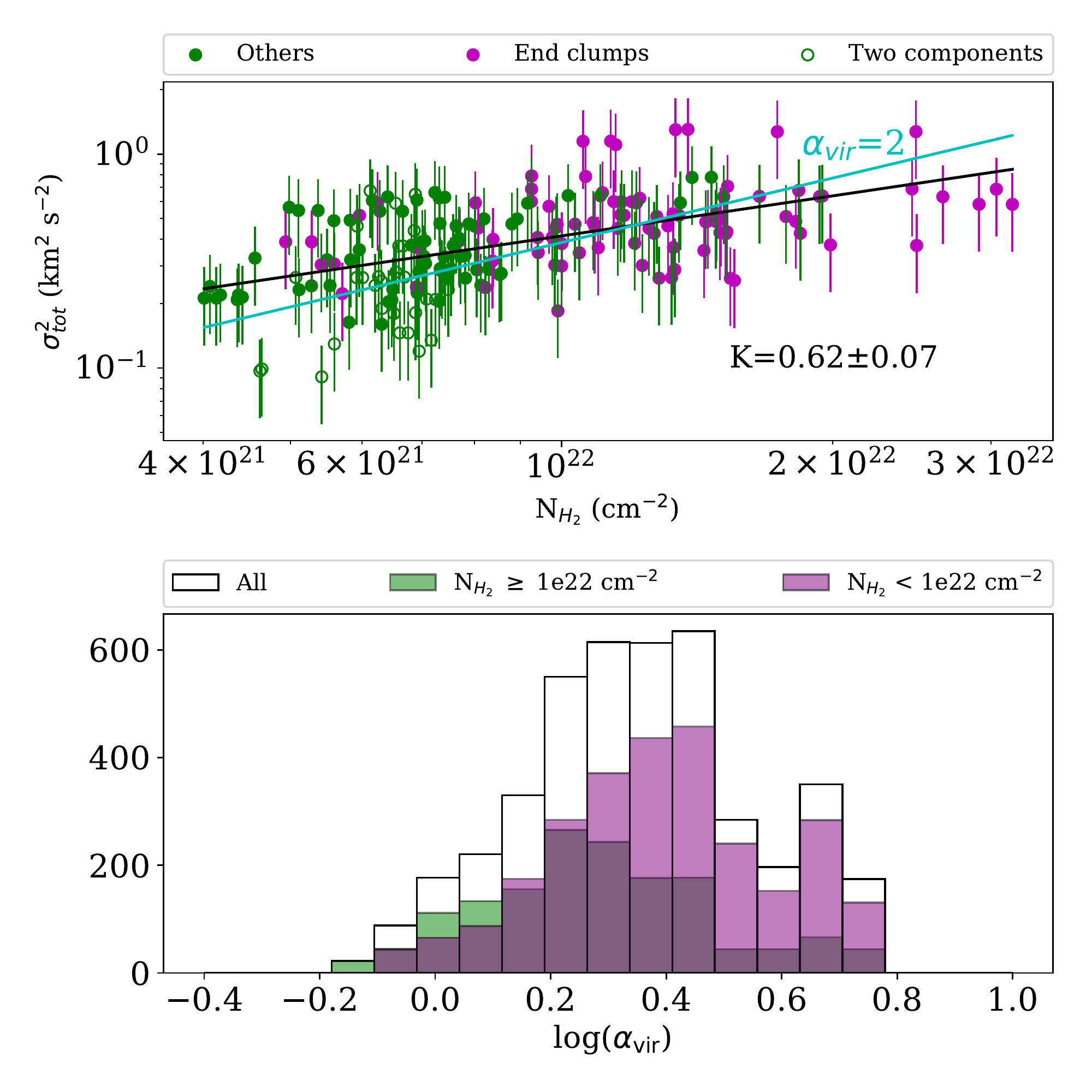}
\caption{\textbf{Upper panel (a)}: Velocity dispersion ($\sigma_{\rm tot}^{2}$) against the the H$_{2}$ column density (N$_{\rm H_{2}}$) for the S242 main skeletons. The magenta dots represent the skeleton structures in both end-clumps, the green dots are other skeleton regions  not located in the end regions,  the open green circles represent the value in the length ranges of $\sim$ 6.5 pc -- 8.5 pc and 13.5 pc -- 15.5 pc, which are calculated by the main one of the two velocity components. The vertical lines on the individual data points show their uncertainties. The black line is the best linear fit for $\sigma_{\rm tot}^{2}$ and N$_{\rm H_{2}}$; the fitted slope parameter (K) is 0.62$\pm$0.07.  The cyan lines show the characteristic of $\alpha_{\rm vir}$ = 2.0. \textbf{Lower panel (b)}: Log-scale of $\alpha_{vir}$ histograms for the S242 main skeleton, the S242 main skeleton with N$_{\rm H_{2}}$ $\geq$ 1.0e22, and the S242 skeleton with N$_{\rm H_{2}}$ $<$ 1.0e22, respectively. All histograms are weighted by the value of H$_{2}$ column density. \label{fig:fvir}}
\end{figure}

\begin{figure*}
\centering
\includegraphics[width=\hsize]{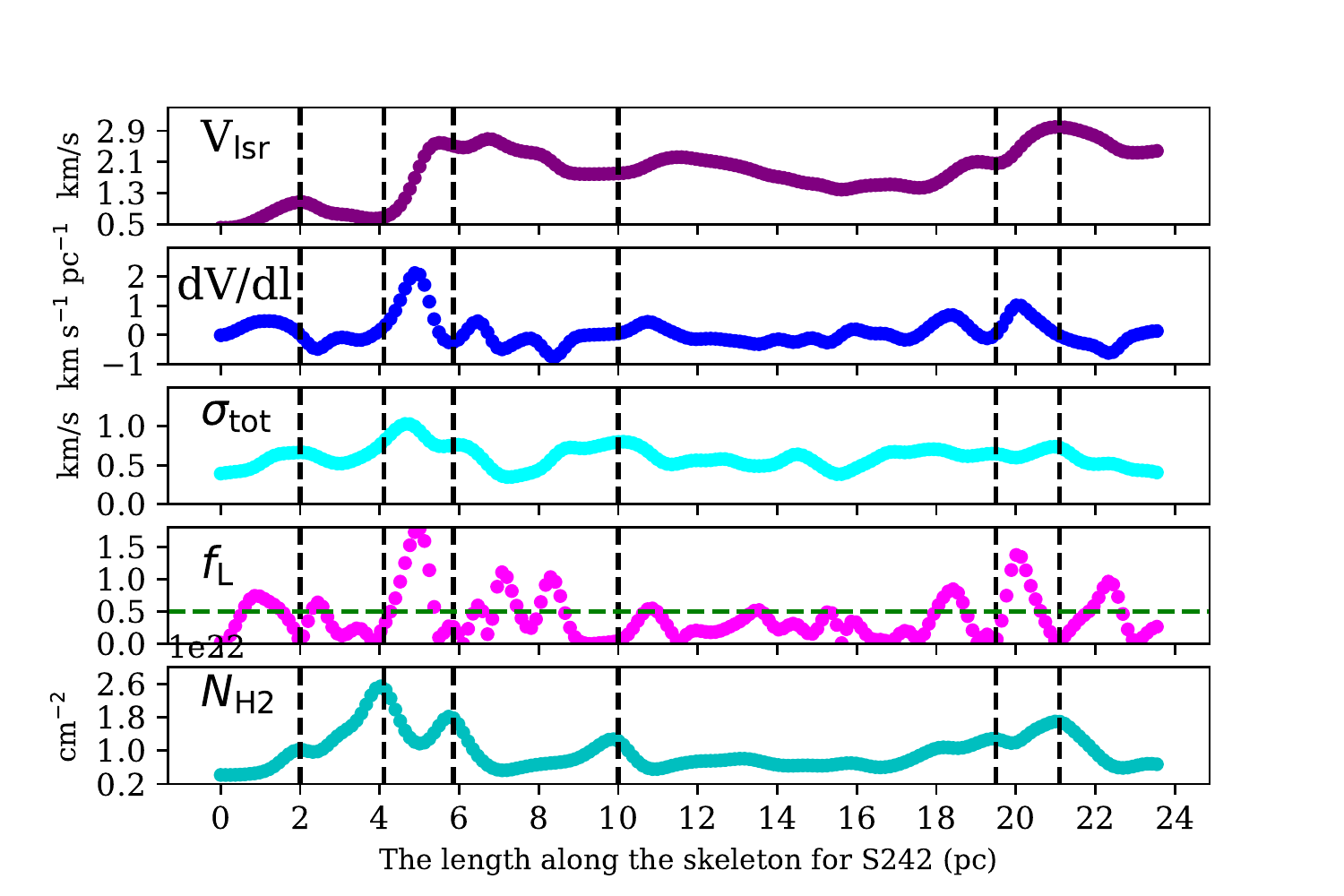}
\caption{Convolved profiles of centroid velocity (V$_{\rm lsr}$), velocity gradients (dV/dl), total 1D velocity dispersion ($\sigma_{\rm tot}$), the contribution of longitudinal collapse in the total velocity dispersion (f$_{\rm L}$), and the H$_{2}$ column density. The vertical black dashed lines indicate the peaks of the H$_{2}$ column density profile; the horizontal green dashed line represents f$_{\rm L}$ = 0.5 \label{fig:flong}}
\end{figure*}

\section{Discussion}
\subsection{Edge collapse of molecular cloud S242}
\subsubsection{The case of the edge collapse in molecular cloud S242}
\cite{Burkert2004}  simulated the collapse and fragmentation in the finite self-gravitating sheets. This simulation shows that the gravitational acceleration has a steep increase at the edges of finite sheets or filaments that causes these structures to collapse immediately with material piling up most rapidly at the outer edges. In addition, the edge-clumps are given more momentum and may further approach the center of filaments \citep{Pon2011, Pon2012}. Moreover, \cite{Clarke2015} suggested that due to the gravitational attraction of the massive end-clumps, the gas ahead of the end-clumps can also be accelerated toward the end-clumps. 

In Figure \ref{fig:fsketch}, we sketch the edge-collapse process of a finite entity. In the time slice of ``Before collapse,'' its geometry is a longer filament.  Due to the gravitational focusing near the closed end of the structure, materials are rapidly accelerated into both  end regions, which results in the formation of gas clumps and star clusters near the ends of structures, as presented in the time slice of ``After collapse'' \citep{Burkert2004, Pon2011, Pon2012}. For the filament S242, the massive clumps and YSOs clusters are mainly located in the end regions \citep{Dewangan2017, Yuanlx2019}.  Moreover, as presented in panel (e) of Figure \ref{fig:fco}, the large-scale velocity gradients distribute along S242; the  blue-shift in the southern end and the red-shift in the northern end may indicate the approaching movement between the two end-clumps. These all suggest that S242 is formed as a result of the global gravity acting on a finite cloud with non-spherical geometry, such as filaments.

\subsubsection{Collapse timescale for the molecular cloud S242}
We illustrate an evolutionary scenario for filament S242 in Figure \ref{fig:fsketch}. The initial filament S242 has a length of L$_{\rm before}$. After the edge collapse, its length becomes L$_{\rm after}$. Considering all the subregions of S242 listed in  Table 1 of \cite{Dewangan2019}, the total mass of filament S242 (M) derived from the C$^{18}$O line, is estimated to be about 9700 M$_{\odot}$. Before collapsing, the line mass of filament S242 (M$_{\rm line}$) is thought to be 209 M$_{\odot}$ pc$^{-1}$, the observed line mass of a subregion between two end regions \citep{Dewangan2019}. Therefore, L$_{\rm before}$= M/M$_{\rm line}$ $\simeq$ 46 pc. The observed filament S242 has a length of $\sim$ 25 pc, and therefore L$_{\rm after}$ $\simeq$ 25 pc. 

To estimate the age of  filament S242 we consider a filament of length L and width d; the gravitational acceleration at one end of the filament can be calculated approximately as
\begin{equation}
    a \simeq \int_{d}^{L}\frac{GM_{\rm line}}{r^{2}} dr \simeq GM_{\rm line} \left(\frac{1}{d}-\frac{1}{L}\right)\ ,
\end{equation}
where G is the gravitational constant, M$_{\rm line}$ is the line mass, and M$_{\rm line}$= dM/dL $\simeq$ 209 M$_{\odot}$ pc$^{-1}$. For filament S242, the length L $\simeq$ 25 pc, the width d $\simeq$ 0.86 pc, and the derived gravitational acceleration is about 1.1 pc Myr$^{-2}$. Under the gravitational acceleration at the end regions, the length of filament S242 becomes L$_{\rm after}$; therefore,
\begin{equation}
\Delta L= L_{\rm before} - L_{\rm after} = 2 dL\ .
\end{equation}
We further estimate the collapse time as
\begin{equation}
t_{\rm col} \simeq \sqrt{2 dL / a} = \sqrt{\Delta L / a}\ .
\end{equation}
The collapse time of filament S242 is about 4.2 Myr; if we take the projection effect into account and assume an inclination ($\theta$) of 30 degrees, the value is about 4.2/$\sqrt{\rm cos(\theta}$) Myr $\simeq$ 4.5 Myr. \cite{Dewangan2019}  have  estimated a collapse timescale of $\sim$ 3.5 Myr for filament S242, which is derived as t$_{\rm col}$=(0.49+0.26A)/(G$\rho$)$^{1/2}$ for the filamentary structures in \cite{Clarke2015}, where A is the initial aspect ratio of filament. They took the observed aspect ratio (A) and a mean density ($\rho$) of the observed S242 filament to calculate the collapse timescale. Their value is the time for the observed filament S242 to collapse into a single dense core and different from the collapse time we estimated; our value is the timescale from the ``Before collapse'' to ``After collapse'' in  Figure \ref{fig:fsketch}. During our timescale, the S242 cloud undergoes the process of edge collapse from its initial structure to the current observed filament. 

In addition, \citet{Yuanlx2019} and \citet{Dewangan2019}  reveal a number of YSO clusters (including Class I and Class II) concentrated mainly in the end clumps. Their average lifetime should be in the range of $\sim$ 0.5 -- 2 Myr \citep{Evans2009}. As shown in panel (e) of Figure 1, the relative velocity of the two end-clumps is about 3 km s$^{-1}$; we take the average velocity (v$_{\rm aver}$) of 1.5 km s$^{-1}$ for the two end-clumps. During the lifetime of Class II YSOs, we estimate that the moving distance for the end-clumps is about 2 Myr $\times$ 1.5 km s$^{-1}$ $\sim$ 3.0 pc. This distance is longer than the size of the dense clumps ($\sim$ 1 pc for substructures in G181) that harbor the YSOs clusters. Therefore, we suggest that the old YSOs previously formed may  also be moving toward the center of the  filament, together with the end-clumps.

\subsection{Gravitational stability and gravity-driven longitudinal accretion in filament S242}
\subsubsection{Self-gravity as the cause of increased velocity dispersion?}
In panel (b) of Figure \ref{fig:fskeleton_profile}, we find a similarity between the $\sigma_{\rm tot}$ profile and that of N$_{\rm H_{2}}$, where the peaks of N$_{\rm H_{2}}$ coincide with the peaks of $\sigma_{\rm tol}$. Furthermore, the peaks of N$_{\rm H_{2}}$ also match the maximum or minimum values of the centroid-velocity profile. Therefore, the local velocity perturbations may be related to the gravitational perturbation caused by the varying line mass. In the upper panel of Figure \ref{fig:fvir}, we plot the velocity dispersion ($\sigma_{\rm tot}^{2}$) against the H$_{2}$ column density (N$_{\rm H_{2}}$) for the main skeleton; we calculate that their Spearman's correlation coefficient is about 0.55, indicating a roughly positive correlation between the H$_{2}$ column density (N$_{\rm H_{2}}$) and velocity dispersion ($\sigma_{\rm tot}^{2}$). The best-fit slope for the linear relationship between log(N$_{\rm H_{2}}$) and log($\sigma_{\rm tot}^{2}$) is 0.62$\pm$0.07. This may imply that the increased velocity dispersion is caused by the matter concentration under self-gravity. Other studies also mentioned that gravitational collapse could generate additional turbulence motion \citep{Klessen2010, Peretto2014, Hacar2017}.

Can the self-gravity account for the increased velocity dispersion? We further estimate the self-gravity induced velocities as  
\begin{equation}
v_{\rm dyn} \simeq \sqrt{\frac{Gm}{r}} \simeq \sqrt{GM_{\rm line}}\ ,
\end{equation}
where G is the gravitational constant; $M_{\rm line}$ is the line mass  estimated as  $M_{\rm line}$= $\mu$$\rm m_{H}N_{H_{2}}$d, where the N$_{\rm H_{2}}$  is the H$_{2}$ column density along the skeleton and d is defined as half of the S242 filament width ( $\sim$ 0.4 pc). Furthermore, for a filament the virial parameters could be calculated as \citep{Ostriker1964, Watkins2019}
\begin{equation}
\alpha_{\rm vir} \simeq \frac{2\sigma_{\rm tot}^{2}}{GM_{\rm line}} \simeq \frac{2\sigma_{\rm tot}^{2}}{v_{\rm dyn}^{2}}\ ,
\end{equation}
where the prefactor of 2 depends on the equation of state and the density profile in filament S242 \citep{Chandrasekhar1953}, which is consistent with the value in \cite{Watkins2019}.

In the upper panel of Figure \ref{fig:fvir}, the cyan line shows the limit for the parameter $\alpha_{\rm vir}$ = 2. The log-scale of $\alpha_{\rm vir}$ histograms for the S242 skeletons are presented in the lower panel of Figure \ref{fig:fvir}, which is weighted by the value of H$_{2}$ column density. We find that the distribution of the log($\alpha_{\rm vir}$) concentrates in the range of $\sim$ 0.2 -- 0.5 (1.6 -- 3.1 for $\alpha_{\rm vir}$). In addition, we also present the distribution of log($\alpha_{\rm vir}$) for the skeleton gas with N$_{\rm H_{2}}$ $\geq$ 1$\times$ 10$^{22}$ and N$_{\rm H_{2}}$ $<$ 1$\times$ 10$^{22}$, respectively. The log($\alpha_{\rm vir}$) in the gas associated with  N$_{\rm H_{2}}$ $\geq$ 1$\times$ 10$^{22}$ mainly distributes in the range of $\sim$ 0.2 -- 0.3 (1.6 -- 2.0 for $\alpha_{\rm vir}$), which is lower than the value of $\sim$ 0.3 -- 0.5 (2.0 -- 3.1 for $\alpha_{\rm vir}$) for the log($\alpha_{\rm vir}$)  in gas corresponding to N$_{\rm H_{2}}$ $<$ 1$\times$ 10$^{22}$. Overall, self-gravity accounts for a higher fraction of velocity dispersion in the regions with high surface mass than that with low surface mass.

It should be noted that the importance of gravity can only be partially reflected by the virial parameter. This is because gravity is balanced by a combination of turbulence motion, radial collapse, and longitudinal collapse. When we evaluate the virial parameters, the velocity dispersion contrails are caused by turbulence, radial collapse, and longitudinal collapse. However, due to the projection, only a small fraction of $\sigma_{\rm longi}$ is contained in our velocity dispersion

\subsubsection{Contribution of velocity dispersions from the longitudinal collapse}

The longitudinal velocity profile for the main skeleton is shown in panel (b) of Figure \ref{fig:fskeleton_profile}. There are  large-scale velocity gradients distributed along filament S242 from the blue-shifted S242-S to red-shifted northern end. How much does the longitudinal collapse contribute to the velocity dispersion? We propose the following equation through which the velocity dispersion of the filament can be decomposed:
\begin{equation}
\rm \sigma_{\rm tot}^{2} = \sigma_{\rm rand}^{2} + \sigma_{\rm radial}^{2} + \sigma_{\rm longi}^{2} \ .
\end{equation}
Here $\sigma_{\rm tot}^{2}$ is the total turbulent motion of gas, which includes $\sigma_{\rm rand}^{2}$, $\sigma_{\rm radial}^{2}$, and $\sigma_{\rm longi}^{2}$, representing the random, radial, and longitudinal gas motion, respectively. 

To estimate the contribution of longitudinal collapse, the profiles of longitudinal velocity (V$_{\rm lsr}$), velocity gradients (dV/dl), velocity dispersion ($\sigma_{\rm tot}$), and H$_{2}$ column density (N$_{\rm H_{2}}$) are convolved to the spatial resolution of the average filament width of 0.86 pc, using a Gaussian kernel and weighting by the value of H$_{2}$ column density. The resultant profiles are presented in Figure \ref{fig:flong}. We further calculate  the contribution of longitudinal velocity gradients to the overall velocity dispersion in a width (d) of 0.86 pc, which is defined as longitudinal factor f$_{\rm L}$ = (dV/dl)$\times$d/$\sigma_{\rm tot}$ = $\sigma_{\rm longi}$/$\sigma_{\rm tot}$. The distribution of longitudinal factor f$_{\rm L}$ along the main skeleton is presented in Figure \ref{fig:flong}; we find that  values of f$_{\rm L}$ greater than 0.5 are mainly located in the structures near the end-clumps, which may be caused by the longitudinal collapse of end-clumps or gravitational attraction by the massive end-clumps. However, the origin of the observed velocity dispersion in substructures with the factor of f$_{\rm L}$ smaller than 0.5 is likely due to the radial collapse or random motion of the cloud.

We note that the value of $\sigma_{\rm longi}$ and f$_{\rm L}$ are affected by the inclination of filament S242. If an inclination ($\theta$) of 30 degrees is assumed, the value of $\sigma_{\rm longi}$ could be underestimated by a factor of 1/sin$(\theta)$ $\sim$ 2. For the structures near the end-clumps, the longitudinal accretion becomes the dominant factor.

\begin{figure*}
\centering
\includegraphics[width=\hsize]{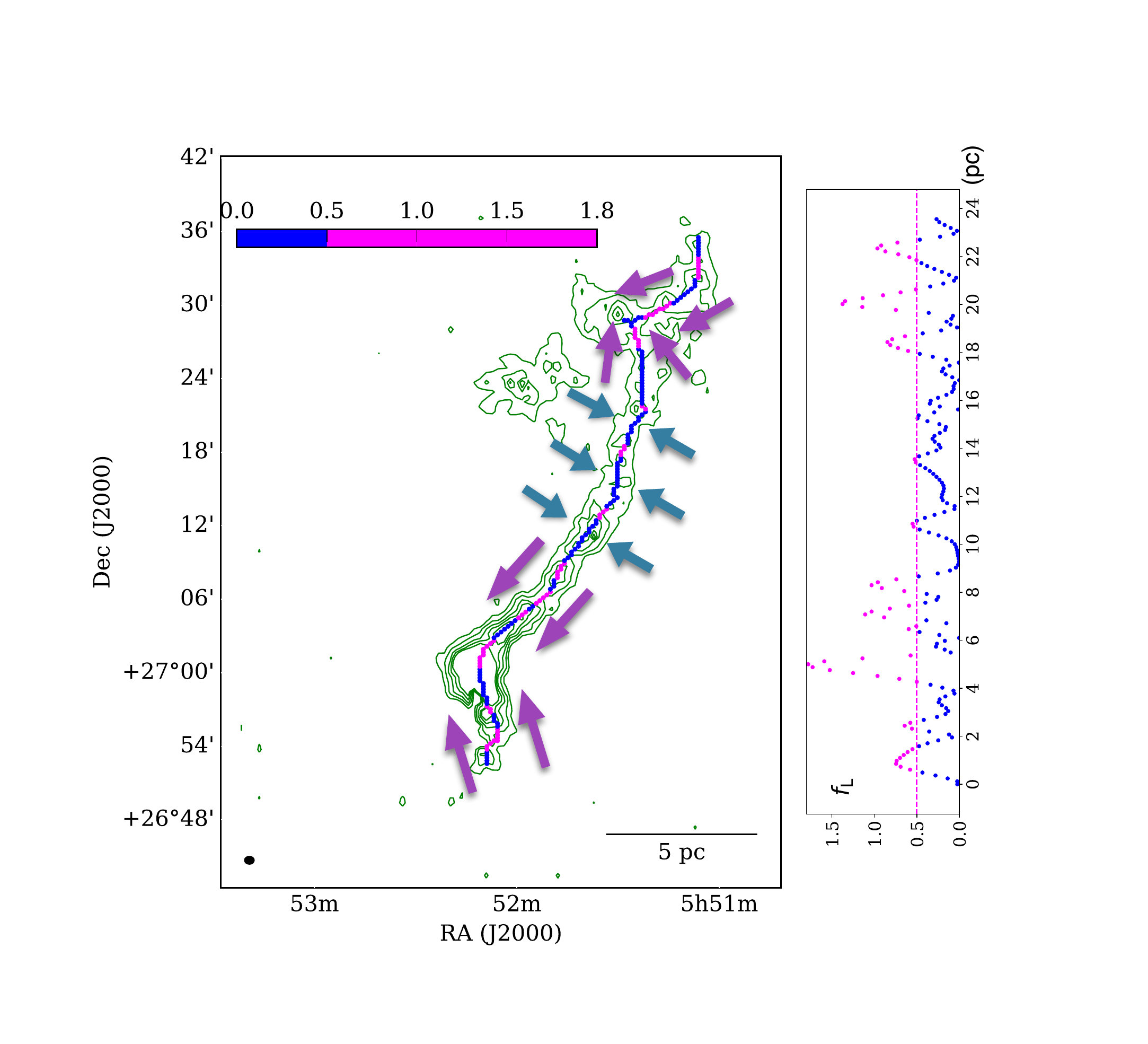}
\caption{Sketch of the subsequent longitudinal accretion of the filament S242. The overlaid green contours display $^{13}$CO column density distribution of filament S242, ranging from 15 \%\ to 90 \%\ stepped by  15 \%\ of the value (1.5 $\times$ 10$^{16}$ cm$^{-2}$). The skeleton structures of the filament S242 are color-coded by the longitudinal factor  f$_{L}$, which is the contribution of longitudinal collapse for the observed velocity dispersion. The magenta arrows indicate the direction of gas motion in longitudinal collapse and accretion, the blue arrows   in   radial collapse. The right subpanel presents the profile of longitudinal factor f$_{\rm L}$ along the spine of filament S242 \label{fig:fsketch_accretion}}
\end{figure*}

\subsubsection{Subsequent longitudinal-accretion caused by the massive end-clumps}
The regions where the   fraction of velocity dispersion from the longitudinal collapse is high are mainly located in the structures near end-clumps. These structures present the higher velocity-gradients (up to $\sim$ 2 km s$^{-1}$ pc$^{-1}$). What is the origin of the longitudinal velocity gradients? The blue-shifted S242-S and red-shifted northern end-clumps imply the approaching movement between the two end-clumps in the edge collapse of filament S242 \citep{Burkert2004, Pon2011, Pon2012}. However, the larger velocity-gradients of $\sim$ 1 km s$^{-1}$ pc$^{-1}$ against the end-clumps may be due to the accretion flows driven by the gravitational attraction of massive end-clumps, as suggested in \cite{Clarke2015}. As shown in Figure \ref{fig:fsketch_accretion}, we sketch the gas motion in the collapse process of filament S242. The magenta sub-skeletons represent the substructures with the value of factor (f$_{\rm L}$) larger than 0.5, which means the longitudinal collapse--accretion dominates the increasing velocity dispersion. The blue regions with the f$_{\rm L}$ smaller than 0.5, which are mainly located in substructures between the two end-clumps, are either under the radial collapse or turbulence motion. Combining with the velocity field for filament S242, we propose the following pictures in Figure \ref{fig:fsketch_accretion}, which indicate the directions of gas motion in the collapse of filament S242.

We suggest that there are two stages in the formation of filament S242. First, the collapse of a finite and elongated filament is affected by edge-effect and the end-clumps acquire a significant amount of mass. Second, the longitudinal accretion becomes dominated toward massive end-clumps, due to their local gravitational attraction.

\section{Conclusions}
The main results of this study are summarized as follows:
   \begin{enumerate}
      \item Filament S242 is regarded as a candidate for  edge collapse,  a mechanism   proposed by \cite{Burkert2004}. The massive clumps and YSOs clusters are mainly located in its end regions \citep{Dewangan2017, Yuanlx2019}. We report the TRAO observations of $^{12}$CO and $^{13}$CO lines for  filament S242. In addition, its distributions of the excitation temperature (8 K -- 25 K), $^{13}$CO column density (0.2 $\times$ 10$^{16}$ cm$^{-2}$ -- 2 $\times$ 10$^{16}$ cm$^{-2}$), line width, and centroid velocity are also presented. We extract the main and coherent velocity component  from the skeleton structures and derive a FWHM of $\sim$ 0.86 pc for filament S242.

      \item We find  a gradual change in velocity along the filament S242. The relative velocity between northern and southern massive end-clumps is about 3 km s$^{-1}$, which may represent the approaching movement between them. These signatures indicate that the filament S242 may start as a finite, truncated filament, where the material is concentrated on the end regions through the edge-collapse mechanism proposed by \cite{Burkert2004}. Moreover, the end-clumps may have acquired more momentum, which causes both end-clumps to approach   the center of the filaments. In addition, we estimate the collapse timescale of filament S242 to be $\sim$ 4.2 Myr, which is the time from the elongated initial structure to the currently observed filament. Compared with the lifetime of YSOs (0.5 Myr -- 2 Myr) and the relative movement of end-clumps, we suggest that the old YSOs are moving toward the center of filament, together with the end-clumps. 
          
      \item We find  that the total velocity dispersion $\sigma_{\rm tot}$ increases correlatively with  the H$_{2}$ column density. In addition, the viral parameters $\alpha_{\rm vir}$ (2$\sigma_{\rm tot}^{2}$/v$_{\rm dyn}^{2}$) in regions with high surface mass tend to be lower than those with low surface mass. Thus, the increased turbulence motion in filament S242 might be gravitationally generated. In addition, we also reveal that the origin of the observed velocity dispersion may be dominated by the longitudinal collapse in the structures in the vicinity of end-clumps; however, for the subfilament between two end-clumps, the velocity dispersion mainly comes from the radial and random motion of the cloud.
      
      \item We propose a two-stage fragmentation scenario for filament S242. First, due to  gravitational focusing, the material in the finite filament is preferentially concentrated at the end regions, and the end-clumps acquire a significant amount of mass. Second, the longitudinal accretion becomes dominant in the vicinity of the high-mass clumps.
   \end{enumerate}

\begin{acknowledgements}
We are grateful to the anonymous referee for a critical reading of the manuscript and constructive suggestions that greatly improved the quality of this paper. Lixia Yuan, Ming Zhu are supported by the National Key R$\&$D Program of China No.2017YFA0402600, the NSFC grant No.U1531246 and also supported by the Open Project Program of the Key Laboratory of FAST, NAOC, Chinese Academy of Sciences. Ke Wang acknowledges support by the National Key Research and Development Program of China (2017YFA0402702, 2019YFA0405100), the National Science Foundation of China (11973013, 11721303), and the starting grant at the Kavli Institute for Astronomy and Astrophysics, Peking University (7101502016). Jinghua Yuan is supported by the NSFC funding No.11503035 and No.11573036. 
\end{acknowledgements}


\bibliographystyle{aa}
\bibliography{reference_G181.bib}

\begin{thebibliography}{65}
\expandafter\ifx\csname natexlab\endcsname\relax\def\natexlab#1{#1}\fi

\bibitem[{{Andr{\'e}}(2017)}]{Andre2017}
{Andr{\'e}}, P. 2017, Comptes Rendus Geoscience, 349, 187

\bibitem[{{Andr{\'e}} {et~al.}(2014){Andr{\'e}}, {Di Francesco},
  {Ward-Thompson}, {Inutsuka}, {Pudritz}, \& {Pineda}}]{Andre2014}
{Andr{\'e}}, P., {Di Francesco}, J., {Ward-Thompson}, D., {et~al.} 2014,
  Protostars and Planets VI, 27

\bibitem[{{Andr{\'e}} {et~al.}(2013){Andr{\'e}}, {K{\"o}nyves}, {Arzoumanian},
  {Palmeirim}, \& {Peretto}}]{Andre2013}
{Andr{\'e}}, P., {K{\"o}nyves}, V., {Arzoumanian}, D., {Palmeirim}, P., \&
  {Peretto}, N. 2013, in Astronomical Society of the Pacific Conference Series,
  Vol. 476, New Trends in Radio Astronomy in the ALMA Era: The 30th Anniversary
  of Nobeyama Radio Observatory, ed. R.~{Kawabe}, N.~{Kuno}, \& S.~{Yamamoto},
  95

\bibitem[{{Andr{\'e}} {et~al.}(2016){Andr{\'e}}, {Rev{\'e}ret}, {K{\"o}nyves},
  {Arzoumanian}, {Tig{\'e}}, {Gallais}, {Roussel}, {Le Pennec}, {Rodriguez},
  {Doumayrou}, {Dubreuil}, {Lortholary}, {Martignac}, {Talvard}, {Delisle},
  {Visticot}, {Dumaye}, {De Breuck}, {Shimajiri}, {Motte}, {Bontemps},
  {Hennemann}, {Zavagno}, {Russeil}, {Schneider}, {Palmeirim}, {Peretto},
  {Hill}, {Minier}, {Roy}, \& {Rygl}}]{Andre2016}
{Andr{\'e}}, P., {Rev{\'e}ret}, V., {K{\"o}nyves}, V., {et~al.} 2016, \aap,
  592, A54

\bibitem[{{Bally} {et~al.}(1987){Bally}, {Langer}, {Stark}, \&
  {Wilson}}]{Bally1987}
{Bally}, J., {Langer}, W.~D., {Stark}, A.~A., \& {Wilson}, R.~W. 1987, \apjl,
  312, L45

\bibitem[{{Binney} \& {Tremaine}(1987)}]{Binney1987}
{Binney}, J. \& {Tremaine}, S. 1987, {Galactic dynamics}

\bibitem[{{Burkert} \& {Hartmann}(2004)}]{Burkert2004}
{Burkert}, A. \& {Hartmann}, L. 2004, \apj, 616, 288

\bibitem[{{Busquet} {et~al.}(2013){Busquet}, {Zhang}, {Palau}, {Liu},
  {S{\'a}nchez-Monge}, {Estalella}, {Ho}, {de Gregorio-Monsalvo}, {Pillai},
  {Wyrowski}, {Girart}, {Santos}, \& {Franco}}]{Busquet2013}
{Busquet}, G., {Zhang}, Q., {Palau}, A., {et~al.} 2013, \apjl, 764, L26

\bibitem[{{Caselli} {et~al.}(1999){Caselli}, {Walmsley}, {Tafalla}, {Dore}, \&
  {Myers}}]{Caselli1999}
{Caselli}, P., {Walmsley}, C.~M., {Tafalla}, M., {Dore}, L., \& {Myers}, P.~C.
  1999, \apj, 523, L165

\bibitem[{{Chandrasekhar} \& {Fermi}(1953)}]{Chandrasekhar1953}
{Chandrasekhar}, S. \& {Fermi}, E. 1953, \apj, 118, 116

\bibitem[{{Clarke} \& {Whitworth}(2015)}]{Clarke2015}
{Clarke}, S.~D. \& {Whitworth}, A.~P. 2015, \mnras, 449, 1819

\bibitem[{{Contreras} {et~al.}(2016){Contreras}, {Garay}, {Rathborne}, \&
  {Sanhueza}}]{Contreras2016}
{Contreras}, Y., {Garay}, G., {Rathborne}, J.~M., \& {Sanhueza}, P. 2016,
  \mnras, 456, 2041

\bibitem[{{Curry}(2000)}]{Curry2000}
{Curry}, C.~L. 2000, The Astrophysical Journal, 541, 831

\bibitem[{{Dewangan} {et~al.}(2017){Dewangan}, {Baug}, {Ojha}, {Janardhan},
  {Devaraj}, \& {Luna}}]{Dewangan2017}
{Dewangan}, L.~K., {Baug}, T., {Ojha}, D.~K., {et~al.} 2017, \apj, 845, 34

\bibitem[{{Dewangan} {et~al.}(2019){Dewangan}, {Pirogov}, {Ryabukhina}, {Ojha},
  \& {Zinchenko}}]{Dewangan2019}
{Dewangan}, L.~K., {Pirogov}, L.~E., {Ryabukhina}, O.~L., {Ojha}, D.~K., \&
  {Zinchenko}, I. 2019, \apj, 877, 1

\bibitem[{{Evans} {et~al.}(2001){Evans}, {Allen}, {Blake}, {Harvey}, {Koerner},
  {Mundy}, {Myers}, {Padgett}, {Sargent}, \& {Stapelfeldt}}]{Evans2001}
{Evans}, N.~J., I., {Allen}, L.~E., {Blake}, G.~A., {et~al.} 2001, in American
  Astronomical Society Meeting Abstracts, Vol. 198, American Astronomical
  Society Meeting Abstracts \#198, 25.05

\bibitem[{{Evans} {et~al.}(2009){Evans}, {Dunham}, {J{\o}rgensen}, {Enoch},
  {Mer{\'{\i}}n}, {van Dishoeck}, {Alcal{\'a}}, {Myers}, {Stapelfeldt},
  {Huard}, {Allen}, {Harvey}, {van Kempen}, {Blake}, {Koerner}, {Mundy},
  {Padgett}, \& {Sargent}}]{Evans2009}
{Evans}, II, N.~J., {Dunham}, M.~M., {J{\o}rgensen}, J.~K., {et~al.} 2009,
  \apjs, 181, 321

\bibitem[{{Fuller} \& {Myers}(1992)}]{Fuller1992}
{Fuller}, G.~A. \& {Myers}, P.~C. 1992, \apj, 384, 523

\bibitem[{Goldsmith {et~al.}(2008)Goldsmith, Heyer, Narayanan, Snell, Li, \&
  Brunt}]{Goldsmith2008}
Goldsmith, P.~F., Heyer, M., Narayanan, G., {et~al.} 2008, The Astrophysical
  Journal, 680, 428

\bibitem[{{Gong} {et~al.}(2018){Gong}, {Li}, {Mao}, {Henkel}, {Menten}, {Fang},
  {Wang}, \& {Sun}}]{Gong2018}
{Gong}, Y., {Li}, G.~X., {Mao}, R.~Q., {et~al.} 2018, \aap, 620, A62

\bibitem[{{Hacar} {et~al.}(2016){Hacar}, {Alves}, {Burkert}, \&
  {Goldsmith}}]{Hacar2016}
{Hacar}, A., {Alves}, J., {Burkert}, A., \& {Goldsmith}, P. 2016, \aap, 591,
  A104

\bibitem[{{Hacar} {et~al.}(2017){Hacar}, {Alves}, {Tafalla}, \&
  {Goicoechea}}]{Hacar2017}
{Hacar}, A., {Alves}, J., {Tafalla}, M., \& {Goicoechea}, J.~R. 2017, \aap,
  602, L2

\bibitem[{{Hacar} {et~al.}(2013){Hacar}, {Tafalla}, {Kauffmann}, \&
  {Kov{\'a}cs}}]{Hacar2013}
{Hacar}, A., {Tafalla}, M., {Kauffmann}, J., \& {Kov{\'a}cs}, A. 2013, \aap,
  554, A55

\bibitem[{{Hartmann} \& {Burkert}(2007)}]{Hartmann2007}
{Hartmann}, L. \& {Burkert}, A. 2007, \apj, 654, 988

\bibitem[{{Heitsch} \& {Hartmann}(2008)}]{Heitsch2008}
{Heitsch}, F. \& {Hartmann}, L. 2008, \apj, 689, 290

\bibitem[{{Hunter} \& {Massey}(1990)}]{Hunter1990}
{Hunter}, D.~A. \& {Massey}, P. 1990, \aj, 99, 846

\bibitem[{{Inutsuka} \& {Miyama}(1992)}]{Inutsuka1992}
{Inutsuka}, S.-I. \& {Miyama}, S.~M. 1992, \apj, 388, 392

\bibitem[{{Johnstone} {et~al.}(2017){Johnstone}, {Ciccone}, {Kirk}, {Mairs},
  {Buckle}, {Berry}, {Broekhoven-Fiene}, {Currie}, {Hatchell}, {Jenness},
  {Mottram}, {Pattle}, {Tisi}, {Di Francesco}, {Hogerheijde}, {Ward-Thompson},
  {Bastien}, {Bresnahan}, {Butner}, {Chen}, {Chrysostomou}, {Coud{\'e}},
  {Davis}, {Drabek-Maunder}, {Duarte-Cabral}, {Fich}, {Fiege}, {Friberg},
  {Friesen}, {Fuller}, {Graves}, {Greaves}, {Gregson}, {Holland}, {Joncas},
  {Kirk}, {Knee}, {Marsh}, {Matthews}, {Moriarty-Schieven}, {Mowat}, {Nutter},
  {Pineda}, {Salji}, {Rawlings}, {Richer}, {Robertson}, {Rosolowsky}, {Rumble},
  {Sadavoy}, {Thomas}, {Tothill}, {Viti}, {White}, {Wouterloot}, {Yates}, \&
  {Zhu}}]{Johnstone2017}
{Johnstone}, D., {Ciccone}, S., {Kirk}, H., {et~al.} 2017, \apj, 836, 132

\bibitem[{{Kainulainen} {et~al.}(2013){Kainulainen}, {Ragan}, {Henning}, \&
  {Stutz}}]{Kainulainen2013}
{Kainulainen}, J., {Ragan}, S.~E., {Henning}, T., \& {Stutz}, A. 2013, \aap,
  557, A120

\bibitem[{{Kainulainen} {et~al.}(2017){Kainulainen}, {Stutz}, {Stanke},
  {Abreu-Vicente}, {Beuther}, {Henning}, {Johnston}, \&
  {Megeath}}]{Kainulainen2017}
{Kainulainen}, J., {Stutz}, A.~M., {Stanke}, T., {et~al.} 2017, \aap, 600, A141

\bibitem[{{Kirk} {et~al.}(2013){Kirk}, {Myers}, {Bourke}, {Gutermuth},
  {Hedden}, \& {Wilson}}]{Kirk2013}
{Kirk}, H., {Myers}, P.~C., {Bourke}, T.~L., {et~al.} 2013, \apj, 766, 115

\bibitem[{{Klessen} \& {Hennebelle}(2010)}]{Klessen2010}
{Klessen}, R.~S. \& {Hennebelle}, P. 2010, \aap, 520, A17

\bibitem[{{Koch} \& {Rosolowsky}(2015)}]{Koch2015}
{Koch}, E.~W. \& {Rosolowsky}, E.~W. 2015, \mnras, 452, 3435

\bibitem[{{Kramer} {et~al.}(1999){Kramer}, {Alves}, {Lada}, {Lada}, {Sievers},
  {Ungerechts}, \& {Walmsley}}]{Kramer1999}
{Kramer}, C., {Alves}, J., {Lada}, C.~J., {et~al.} 1999, \aap, 342, 257

\bibitem[{{Larson}(1985)}]{Larson1985}
{Larson}, R.~B. 1985, \mnras, 214, 379

\bibitem[{{Li} {et~al.}(2016{\natexlab{a}}){Li}, {Urquhart}, {Leurini},
  {Csengeri}, {Wyrowski}, {Menten}, \& {Schuller}}]{Ligx2016}
{Li}, G.-X., {Urquhart}, J.~S., {Leurini}, S., {et~al.} 2016{\natexlab{a}},
  \aap, 591, A5

\bibitem[{{Li} {et~al.}(2013){Li}, {Wyrowski}, {Menten}, \&
  {Belloche}}]{Ligx2013}
{Li}, G.-X., {Wyrowski}, F., {Menten}, K., \& {Belloche}, A. 2013, \aap, 559,
  A34

\bibitem[{{Li} {et~al.}(2016{\natexlab{b}}){Li}, {Klein}, \& {McKee}}]{Li2016}
{Li}, P.~S., {Klein}, R.~I., \& {McKee}, C.~F. 2016{\natexlab{b}}, in IAU
  Symposium, Vol. 315, From Interstellar Clouds to Star-Forming Galaxies:
  Universal Processes?, ed. P.~{Jablonka}, P.~{Andr{\'e}}, \& F.~{van der Tak},
  103--106

\bibitem[{{Liu} {et~al.}(2018{\natexlab{a}}){Liu}, {Kim}, {Juvela}, {Wang},
  {Tatematsu}, {Di Francesco}, {Liu}, {Wu}, {Thompson}, {Fuller}, {Eden}, {Li},
  {Ristorcelli}, {Kang}, {Lin}, {Johnstone}, {He}, {Koch}, {Sanhueza}, {Qin},
  {Zhang}, {Hirano}, {Goldsmith}, {Evans}, {White}, {Choi}, {Lee}, {Toth},
  {Mairs}, {Yi}, {Tang}, {Soam}, {Peretto}, {Samal}, {Fich}, {Parsons}, {Yuan},
  {Zhang}, {Malinen}, {Bendo}, {Rivera-Ingraham}, {Liu}, {Wouterloot}, {Li},
  {Qian}, {Rawlings}, {Rawlings}, {Feng}, {Aikawa}, {Akhter}, {Alina}, {Bell},
  {Bernard}, {Blain}, {B{\H o}gner}, {Bronfman}, {Byun}, {Chapman}, {Chen},
  {Chen}, {Chen}, {Chen}, {Chen}, {Chrysostomou}, {Cosentino}, {Cunningham},
  {Demyk}, {Drabek-Maunder}, {Doi}, {Eswaraiah}, {Falgarone}, {Feh{\'e}r},
  {Fraser}, {Friberg}, {Garay}, {Ge}, {Gear}, {Greaves}, {Guan},
  {Harvey-Smith}, {HASEGAWA}, {Hatchell}, {He}, {Henkel}, {Hirota}, {Holland},
  {Hughes}, {Jarken}, {Ji}, {Jimenez-Serra}, {Kang}, {Kawabata}, {Kim}, {Kim},
  {Kim}, {Kim}, {Koo}, {Kwon}, {Kuan}, {Lacaille}, {Lai}, {Lee}, {Lee}, {Lee},
  {Li}, {Li}, {Lo}, {Lopez}, {Lu}, {Lyo}, {Mardones}, {Marston}, {McGehee},
  {Meng}, {Montier}, {Montillaud}, {Moore}, {Morata}, {Moriarty-Schieven},
  {Ohashi}, {Pak}, {Park}, {Paladini}, {Pattle}, {Pech}, {Pelkonen}, {Qiu},
  {Ren}, {Richer}, {Saito}, {Sakai}, {Shang}, {Shinnaga}, {Stamatellos},
  {Tang}, {Traficante}, {Vastel}, {Viti}, {Walsh}, {Wang}, {Wang}, {Wang},
  {Ward-Thompson}, {Whitworth}, {Xu}, {Yang}, {Yang}, {Yuan}, {Zavagno},
  {Zhang}, {Zhang}, {Zhou}, {Zhou}, {Zhu}, {Zuo}, \& {Zhang}}]{Liua2018}
{Liu}, T., {Kim}, K.-T., {Juvela}, M., {et~al.} 2018{\natexlab{a}}, \apjs, 234,
  28

\bibitem[{{Liu} {et~al.}(2018{\natexlab{b}}){Liu}, {Li}, {Juvela}, {Kim},
  {Evans}, {Di Francesco}, {Liu}, {Yuan}, {Tatematsu}, {Zhang},
  {Ward-Thompson}, {Fuller}, {Goldsmith}, {Koch}, {Sanhueza}, {Ristorcelli},
  {Kang}, {Chen}, {Hirano}, {Wu}, {Sokolov}, {Lee}, {White}, {Wang}, {Eden},
  {Li}, {Thompson}, {Pattle}, {Soam}, {Nasedkin}, {Kim}, {Kim}, {Lai}, {Park},
  {Qiu}, {Zhang}, {Alina}, {Eswaraiah}, {Falgarone}, {Fich}, {Greaves}, {Gu},
  {Kwon}, {Li}, {Malinen}, {Montier}, {Parsons}, {Qin}, {Rawlings}, {Ren},
  {Tang}, {Tang}, {Toth}, {Wang}, {Wouterloot}, {Yi}, \& {Zhang}}]{Liub2018}
{Liu}, T., {Li}, P.~S., {Juvela}, M., {et~al.} 2018{\natexlab{b}}, \apj, 859,
  151

\bibitem[{{Lu} {et~al.}(2018){Lu}, {Zhang}, {Liu}, {Sanhueza}, {Tatematsu},
  {Feng}, {Smith}, {Myers}, {Sridharan}, \& {Gu}}]{Lu2018}
{Lu}, X., {Zhang}, Q., {Liu}, H.~B., {et~al.} 2018, \apj, 855, 9

\bibitem[{{Mattern} {et~al.}(2018){Mattern}, {Kauffmann}, {Csengeri},
  {Urquhart}, {Leurini}, {Wyrowski}, {Giannetti}, {Barnes}, {Beuther},
  {Bronfman}, {Duarte-Cabral}, {Henning}, {Kainulainen}, {Menten}, {Schisano},
  \& {Schuller}}]{Mattern2018}
{Mattern}, M., {Kauffmann}, J., {Csengeri}, T., {et~al.} 2018, \aap, 619, A166

\bibitem[{{Molinari} {et~al.}(2010){Molinari}, {Swinyard}, {Bally}, {Barlow},
  {Bernard}, {Martin}, {Moore}, {Noriega-Crespo}, {Plume}, {Testi}, {Zavagno},
  {Abergel}, {Ali}, {Anderson}, {Andr{\'e}}, {Baluteau}, {Battersby},
  {Beltr{\'a}n}, {Benedettini}, {Billot}, {Blommaert}, {Bontemps}, {Boulanger},
  {Brand}, {Brunt}, {Burton}, {Calzoletti}, {Carey}, {Caselli}, {Cesaroni},
  {Cernicharo}, {Chakrabarti}, {Chrysostomou}, {Cohen}, {Compiegne}, {de
  Bernardis}, {de Gasperis}, {di Giorgio}, {Elia}, {Faustini}, {Flagey},
  {Fukui}, {Fuller}, {Ganga}, {Garcia-Lario}, {Glenn}, {Goldsmith}, {Griffin},
  {Hoare}, {Huang}, {Ikhenaode}, {Joblin}, {Joncas}, {Juvela}, {Kirk},
  {Lagache}, {Li}, {Lim}, {Lord}, {Marengo}, {Marshall}, {Masi}, {Massi},
  {Matsuura}, {Minier}, {Miville-Desch{\^e}nes}, {Montier}, {Morgan}, {Motte},
  {Mottram}, {M{\"u}ller}, {Natoli}, {Neves}, {Olmi}, {Paladini}, {Paradis},
  {Parsons}, {Peretto}, {Pestalozzi}, {Pezzuto}, {Piacentini}, {Piazzo},
  {Polychroni}, {Pomar{\`e}s}, {Popescu}, {Reach}, {Ristorcelli}, {Robitaille},
  {Robitaille}, {Rod{\'o}n}, {Roy}, {Royer}, {Russeil}, {Saraceno}, {Sauvage},
  {Schilke}, {Schisano}, {Schneider}, {Schuller}, {Schulz}, {Sibthorpe},
  {Smith}, {Smith}, {Spinoglio}, {Stamatellos}, {Strafella}, {Stringfellow},
  {Sturm}, {Taylor}, {Thompson}, {Traficante}, {Tuffs}, {Umana}, {Valenziano},
  {Vavrek}, {Veneziani}, {Viti}, {Waelkens}, {Ward-Thompson}, {White},
  {Wilcock}, {Wyrowski}, {Yorke}, \& {Zhang}}]{Molinari2010}
{Molinari}, S., {Swinyard}, B., {Bally}, J., {et~al.} 2010, \aap, 518, L100

\bibitem[{{Myers}(2009)}]{Myers2009}
{Myers}, P.~C. 2009, \apj, 700, 1609

\bibitem[{{Ohashi} {et~al.}(2018){Ohashi}, {Sanhueza}, {Sakai}, {Kandori},
  {Choi}, {Hirota}, {Nguy{\#7877}n-Lu'o'ng}, \& {Tatematsu}}]{Ohashi2018}
{Ohashi}, S., {Sanhueza}, P., {Sakai}, N., {et~al.} 2018, \apj, 856, 147

\bibitem[{{Ostriker}(1964)}]{Ostriker1964}
{Ostriker}, J. 1964, \apj, 140, 1056

\bibitem[{{Peretto} {et~al.}(2014){Peretto}, {Fuller}, {Andr{\'e}},
  {Arzoumanian}, {Rivilla}, {Bardeau}, {Duarte Puertas}, {Guzman Fernandez},
  {Lenfestey}, {Li}, {Olguin}, {R{\"o}ck}, {de Villiers}, \&
  {Williams}}]{Peretto2014}
{Peretto}, N., {Fuller}, G.~A., {Andr{\'e}}, P., {et~al.} 2014, \aap, 561, A83

\bibitem[{{Pineda} {et~al.}(2010){Pineda}, {Goodman}, {Arce}, {Caselli},
  {Foster}, {Myers}, \& {Rosolowsky}}]{Pineda2010}
{Pineda}, J.~E., {Goodman}, A.~A., {Arce}, H.~G., {et~al.} 2010, \apjl, 712,
  L116

\bibitem[{{Planck Collaboration} {et~al.}(2016){Planck Collaboration}, {Ade},
  {Aghanim}, {Arnaud}, {Ashdown}, {Aumont}, {Baccigalupi}, {Banday},
  {Barreiro}, \& {Bartolo}}]{Planck2016}
{Planck Collaboration}, {Ade}, P.~A.~R., {Aghanim}, N., {et~al.} 2016, \aap,
  594, A28

\bibitem[{{Pon} {et~al.}(2011){Pon}, {Johnstone}, \& {Heitsch}}]{Pon2011}
{Pon}, A., {Johnstone}, D., \& {Heitsch}, F. 2011, \apj, 740, 88

\bibitem[{{Pon} {et~al.}(2012){Pon}, {Toal{\'a}}, {Johnstone},
  {V{\'a}zquez-Semadeni}, {Heitsch}, \& {G{\'o}mez}}]{Pon2012}
{Pon}, A., {Toal{\'a}}, J.~A., {Johnstone}, D., {et~al.} 2012, \apj, 756, 145

\bibitem[{{Qian} {et~al.}(2012){Qian}, {Li}, \& {Goldsmith}}]{Qian2012}
{Qian}, L., {Li}, D., \& {Goldsmith}, P.~F. 2012, \apj, 760, 147

\bibitem[{{Schneider} {et~al.}(2012){Schneider}, {Csengeri}, {Hennemann},
  {Motte}, {Didelon}, {Federrath}, {Bontemps}, {Di Francesco}, {Arzoumanian},
  {Minier}, {Andr{\'e}}, {Hill}, {Zavagno}, {Nguyen-Luong}, {Attard},
  {Bernard}, {Elia}, {Fallscheer}, {Griffin}, {Kirk}, {Klessen}, {K{\"o}nyves},
  {Martin}, {Men'shchikov}, {Palmeirim}, {Peretto}, {Pestalozzi}, {Russeil},
  {Sadavoy}, {Sousbie}, {Testi}, {Tremblin}, {Ward-Thompson}, \&
  {White}}]{Schneider2012}
{Schneider}, N., {Csengeri}, T., {Hennemann}, M., {et~al.} 2012, \aap, 540, L11

\bibitem[{{Schneider} \& {Elmegreen}(1979)}]{Schneider1979}
{Schneider}, S. \& {Elmegreen}, B.~G. 1979, \apjs, 41, 87

\bibitem[{{Schuller} {et~al.}(2009){Schuller}, {Menten}, {Contreras},
  {Wyrowski}, {Schilke}, {Bronfman}, {Henning}, {Walmsley}, {Beuther},
  {Bontemps}, {Cesaroni}, {Deharveng}, {Garay}, {Herpin}, {Lefloch}, {Linz},
  {Mardones}, {Minier}, {Molinari}, {Motte}, {Nyman}, {Reveret}, {Risacher},
  {Russeil}, {Schneider}, {Testi}, {Troost}, {Vasyunina}, {Wienen}, {Zavagno},
  {Kovacs}, {Kreysa}, {Siringo}, \& {Wei{\ss}}}]{Schuller2009}
{Schuller}, F., {Menten}, K.~M., {Contreras}, Y., {et~al.} 2009, \aap, 504, 415

\bibitem[{{Su} {et~al.}(2019){Su}, {Yang}, {Zhang}, {Gong}, {Wang}, {Zhou},
  {Wang}, {Chen}, {Sun}, {Chen}, {Xu}, \& {Jiang}}]{Yang2019}
{Su}, Y., {Yang}, J., {Zhang}, S., {et~al.} 2019, \apjs, 240, 9

\bibitem[{{Tafalla} {et~al.}(2002){Tafalla}, {Myers}, {Caselli}, {Walmsley}, \&
  {Comito}}]{Tafalla2002}
{Tafalla}, M., {Myers}, P.~C., {Caselli}, P., {Walmsley}, C.~M., \& {Comito},
  C. 2002, \apj, 569, 815

\bibitem[{{van Dishoeck} \& {Black}(1988)}]{Van1988}
{van Dishoeck}, E.~F. \& {Black}, J.~H. 1988, \apj, 334, 771

\bibitem[{{Visser} {et~al.}(2009){Visser}, {van Dishoeck}, \&
  {Black}}]{Visser2009}
{Visser}, R., {van Dishoeck}, E.~F., \& {Black}, J.~H. 2009, \aap, 503, 323

\bibitem[{{Wang} {et~al.}(2016){Wang}, {Testi}, {Burkert}, {Walmsley},
  {Beuther}, \& {Henning}}]{Wang2016}
{Wang}, K., {Testi}, L., {Burkert}, A., {et~al.} 2016, \apjs, 226, 9

\bibitem[{{Wang} {et~al.}(2015){Wang}, {Testi}, {Ginsburg}, {Walmsley},
  {Molinari}, \& {Schisano}}]{Wang2015}
{Wang}, K., {Testi}, L., {Ginsburg}, A., {et~al.} 2015, \mnras, 450, 4043

\bibitem[{{Watkins} {et~al.}(2019){Watkins}, {Peretto}, {Marsh}, \&
  {Fuller}}]{Watkins2019}
{Watkins}, E.~J., {Peretto}, N., {Marsh}, K., \& {Fuller}, G.~A. 2019, \aap,
  628, A21

\bibitem[{{Wiseman} \& {Ho}(1998)}]{Wiseman1998}
{Wiseman}, J.~J. \& {Ho}, P.~T.~P. 1998, \apj, 502, 676

\bibitem[{{Yuan} {et~al.}(2019){Yuan}, {Zhu}, {Liu}, {Yuan}, {Wu}, {Kim},
  {Wang}, {Zhou}, {Tatematsu}, \& {Kuno}}]{Yuanlx2019}
{Yuan}, L., {Zhu}, M., {Liu}, T., {et~al.} 2019, \mnras, 487, 1315

\bibitem[{{Zucker} \& {Chen}(2018)}]{Zucker2018}
{Zucker}, C. \& {Chen}, H. H.-H. 2018, \apj, 864, 152

\end{thebibliography}

\begin{appendix} 
\section{ }

  \begin{figure*}
   \centering
     \includegraphics[width=\hsize]{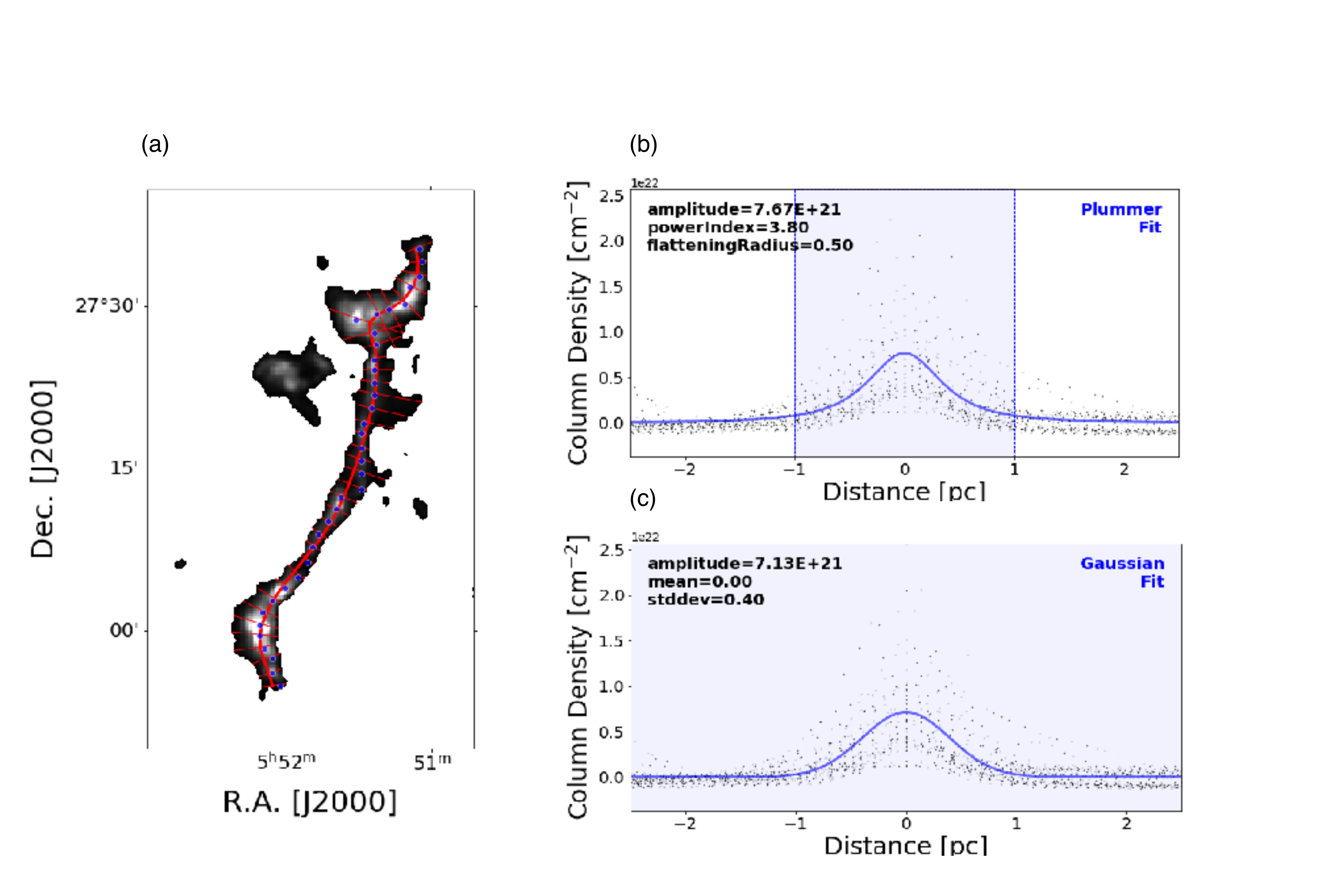}
      \caption{Radial profile for H$_{2}$ column density for the filament S242. \textbf{Panel (a):} Distribution of radial cuts along the spine of filament S242. The red curve represents the spine across the entire filament S242, extracted by FilFinder. The blue dots show the sampling points with an interval of $\sim$ 0.7 pc. The red radial lines,  perpendicular to the tangent of the spine, are the cuts used to calculated the radial distances from the spine. \textbf{Panels (b) and (c):}  Plummer and Gaussian fitting for the radial profiles. The black dots represent the value of N$_{\rm H_{2}}$  along the cuts with respect to their distance from the spine. The Plummer and Gaussian fit is indicated by the blue lines in  Panel (b) and Panel (c), respectively. The best-fit parameters for each function are given in the upper corner of each panel.}
         \label{fwid1}
   \end{figure*}
 
    \begin{figure*}
   \centering
           \includegraphics[width=\hsize]{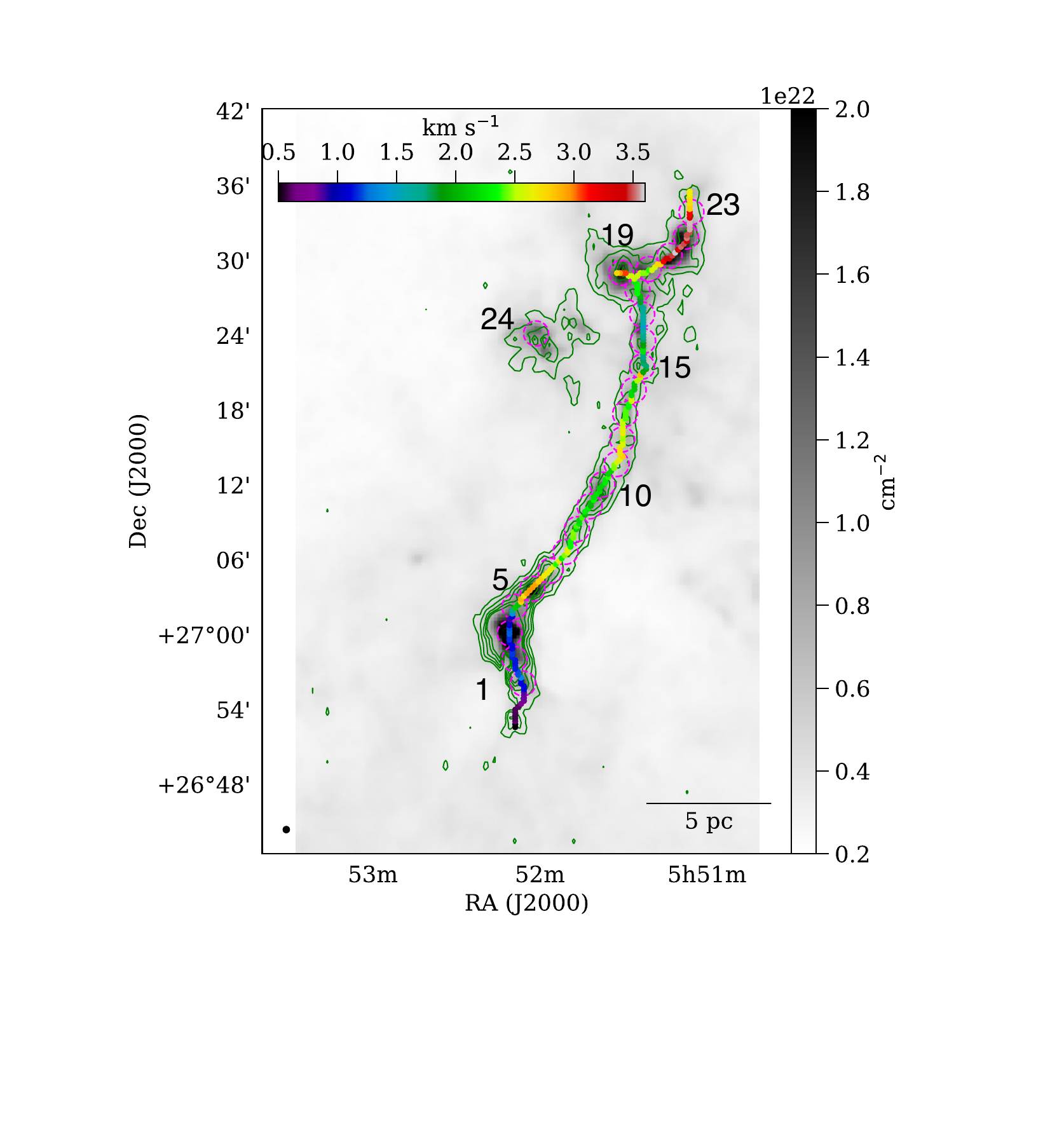}
      \caption{As   in Figure \ref{fig:fskeleton}, but   the magenta circles show the subregions for the average spectra from No.1 to No.24 displayed in Figure A.3.} 
         \label{fspec_reg}
   \end{figure*}

   \begin{figure*}
   \centering
      \includegraphics[width=\hsize]{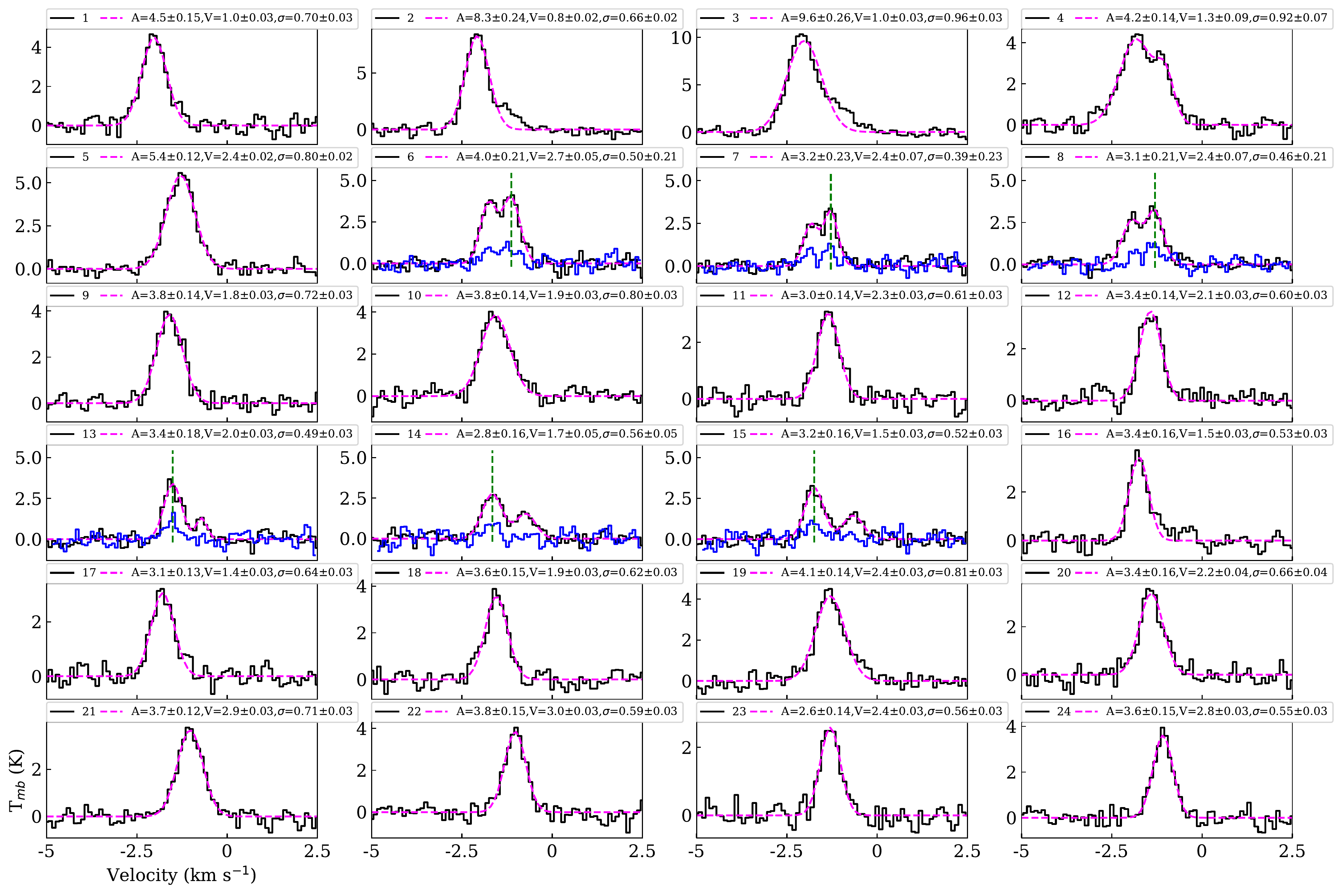}
      \caption{Average spectral profiles for $^{13}$CO lines in the corresponding subregions from No.1 to No.24, displayed in Figure A.2. The black profiles represent the observed spectra, the magenta dashed profiles are   our fitted results by one or two Gaussians. The fitted parameters   for the main velocity components (including peak intensity, centroid velocity, and velocity dispersion) are given  at the top of every subpanel. The blue profiles in No.6 -- No.8 and No.13 -- No.15 are the average C$^{18}$O line spectra, which are amplified by a factor of 3. These C$^{18}$O data are from MWISP CO surveys \citep{Yang2019}. In each of those panels, the vertical green dashed line shows the centroid velocity of the main velocity component.}
         \label{fspec}
   \end{figure*}
\end{appendix}

\end{document}